\documentclass[fleqn,usenatbib]{mnras}

\usepackage{newtxtext,newtxmath}
\usepackage[T1]{fontenc}

\DeclareRobustCommand{\VAN}[3]{#2}
\let\VANthebibliography\thebibliography
\def\thebibliography{\DeclareRobustCommand{\VAN}[3]{##3}\VANthebibliography}

\usepackage{graphicx}	% Including figure files
\usepackage{amsmath}	% Advanced maths commands
\usepackage{pifont}
\usepackage{multirow}
\DeclareRobustCommand{\ion}[2]{%
\relax\ifmmode
\ifx\testbx\f@series
{\mathbf{#1\,\mathsc{#2}}}\else
{\mathrm{#1\,\mathsc{#2}}}\fi
\else\textup{#1\,{\mdseries\textsc{#2}}}%
\fi}

\title[AGN jets in galaxy mergers]{Simulations of spin-driven AGN jets in gas-rich galaxy mergers}

\author[R. Y. Talbot et al.]{
Rosie Y. Talbot$^{1,2,3}$\thanks{E-mail: rosie@mpa-garching.mpg.de (RYT)}, Debora Sijacki$^{2,3}$ and Martin A. Bourne$^{2,3}$
\\
$^{1}$Max-Planck-Institut f{\"u}r Astrophysik, Karl-Schwarzschild-Str. 1, D-85748 Garching, Germany\\
$^{2}$Institute of Astronomy, University of Cambridge, Madingley Road, Cambridge, CB3 0HA, UK\\
$^{3}$Kavli Institute for Cosmology, University of Cambridge, Madingley Road, Cambridge, CB3 0HA, UK
}

\date{Submitted to MNRAS}
\pubyear{2023}

\begin{document}
\label{firstpage}
\pagerange{\pageref{firstpage}--\pageref{lastpage}}
\maketitle

\begin{abstract}
In this work, we use hydrodynamical simulations to explore the effects of kinetic AGN jet feedback on the progression and outcome of the major merger of two isolated, gas-rich galaxies. We present simulations that use the moving-mesh code \textsc{arepo} to follow the progression of the merger through first passage and up to the final coalescence, modelling the black holes at the centres of both of the merging galaxies using our prescription for black hole accretion via an $\alpha$-disc and feedback in the form of a spin-driven jet. We find that the jets drive large-scale, multiphase outflows which launch large quantities of cold gas out to distances greater than $100$~kpc and with velocities that reach $\sim 2500 \, {\rm km \, s^{-1}}$. Gas in the outflows that decelerates, cools and falls back on the galaxies can provide a rich source of fuel for the black hole, leading to intense episodes of jet activity in which the jet can become significantly misaligned. The presence of AGN jets affects the growth of the stellar component: star formation is moderately suppressed at all times during the merger and the peak of the star formation rate, attained during the final coalescence of the galaxies, is reduced by a factor of $\sim 2$. Analysis of simulations such as these will play a central role in making precise predictions for multimessenger investigations of dual radio-AGN, which next-generation observational facilities such as LISA, Athena and SKA will make possible.
\end{abstract}


%%%%%%%%%%%%%%%%%%%%%%%%%%%%%%%%%%%%%%%%%%%%%%%%%%

%%%%%%%%%%%%%%%%% BODY OF PAPER %%%%%%%%%%%%%%%%%%

%%%%%%%%%%%%%%%%%%%%%%%%%%%%%%%%%%%%%%%%%%%%%%%%%%

\section{Introduction}

Galaxy mergers are a natural prediction of hierarchical models of structure formation \citep{1978WhiteRees} and, therefore, play a crucial role in determining the properties of the observed galaxy population. The dynamics of galaxy mergers was first\footnote{Although, see also \citet{1940Holmberg,1941Holmberg} for an ingenious method of simulating galaxy interactions using light-bulbs.} investigated numerically by \citet{1972ToomreToomre} who showed, using restricted N-body techniques, that stellar bridges and tails, as observed in interacting and peculiar galaxies, can form through the action of gravitational tidal torques. Subsequent work, using full N-body methods, confirmed these conclusions \citep{1988Barnes,1992Barnes,1992Hernquist,1993Hernquist}, giving a fairly complete picture of the dynamics of the collisionless component in galaxy mergers.

\citet{1972ToomreToomre}, despite focusing on the collisionless dynamics of mergers, also noted that the observed galaxies they were attempting to model displayed particularly high levels of star formation. They anticipated that mergers could `bring {\it deep} into a galaxy a fairly {\it sudden} supply of fresh fuel in the form of interstellar material'. Indeed, the importance of the gaseous component in merger scenarios is now well established. Numerical simulations have shown that during mergers, extreme tidal torques experienced by gas that cools radiatively can drive significant quantities of gas towards the central regions of the galaxies \citep{1991BarnesHernquist,1996BarnesHernquist}. This abundance of cold gas can lead to highly elevated star formation rates (SFRs) and can also provide the fuel required to power highly energetic outbursts from active galactic nuclei (AGN).

Idealised numerical simulations, including both gaseous and collisionless components, have played a fundamental role in galaxy merger studies over the past few decades. Such simulations have been used to explore topics such as the feeding of and feedback from supermassive black holes \citep[SMBHs; see e.g.][]{2005DiMatteo+,2005Springel+,2009Johansson+,2014Barai+,2014Choi+}, the processes responsible for driving starbursts \citep[see e.g.][]{1996MihosHernquist,2007DiMatteo+,2008DiMatteo+,2008Cox+} and the properties of the merger remnant \citep[see e.g.][]{2003NaabBurkert,2006Cox+, 2009DiMatteo+,2010Wuyts+,2014Perret+}, amongst many others.

In the past decade or so, significant improvements have been made to the numerical models used in such simulations, increasing their physical accuracy and allowing more complex processes to be included. For example, recent idealised merger simulations are now able to include explicit models of stellar feedback alongside detailed models of the interstellar medium (ISM) that capture its multiphase structure \citep{2013Hopkins+,2021Moreno+,2022Li+}. 

AGN feedback is believed to operate in, perhaps, two distinct modes, taking the form of either massive, wide-angle winds or highly-energetic, relativistic jets \citep[see e.g.,][for reviews of AGN feedback]{2007NulsenMcNamara, 2012Fabian, 2015kingPounds, 2018Harrison+}. Recent observations have also shown that ionised outflows may, in fact, be more prevalent and have a greater effect on the surrounding ISM in cases where the radio emission is compact \citep{2015Harrison+,2015Morganti+}. Such powerful expulsions of energy {\it during} the merger process clearly have the potential to dramatically alter the properties of the galaxies before coalescence as well as those of the final remnant and its surroundings. Indeed, numerical simulations of idealised mergers have shown that black hole feedback in the form of AGN winds produces elliptical galaxies that are in much better agreement with observations \citep[for seminal papers on this, see][]{2005DiMatteo+,2005Springel+,2005Springelb+}.

There is now a significant body of simulation work that explores the evolution and effects of AGN jets in idealised galaxy cluster setups \citep{2017BourneSijacki,2017Weinberger+,2018Ehlert+,2019Beckmann+, 2021Su+} and also in disc galaxies, including the interactions of these jets with the ISM \citep{2011WagnerBicknell, 2012Gaibler+, 2016Mukherjee+, 2016Bieri+,2021Talbot+,2022Talbot+}. The effects of kinetic AGN jet feedback in merging galaxies, however, remains largely unexplored.

Galaxy mergers are expected to lead to the formation of SMBH binaries. The low-frequency gravitational waves emitted during the inspiral, merger and ringdown of such binaries is one of the key targets of the LISA mission \citep{2017Amaro-Seoane+} which is sensitive to the coalescence of SMBHs in the mass range $10^{4-7}\, {\rm M_\odot}$ all the way out to $z \approx 20$ \citep{2019Colpi+}. Detecting merging SMBHs that host discernible radio jets is incredibly difficult and, to date, only a few dual radio AGN candidates have been observed \citep[see e.g.][]{2017Kharb+,2017Bansal+,2018Britzen+,2021Dey+,2022MGopal-Krishna+}. Large-scale surveys in the radio, carried out by upcoming observatories such as the Next Generation Very Large Array (ngVLA) and the Square Kilometre Array (SKA), however, will significantly increase the number of dual AGN observed at sub-kpc separations \citep{2015Paragi+}. Data from these instruments will also provide new insights into the nature and properties of AGN jets, particularly at low frequencies. With X-ray missions such as Athena and Lynx on the horizon, as well as the newly operational JWST, the multimessenger study of merging SMBHs will become a reality.

Using numerical simulations of galaxy mergers to investigate the formation and evolution of dual, jetted AGN and their effect on their surroundings is, therefore, exceptionally timely. Such simulations will play a central role in providing firm theoretical predictions for, as well as accurately interpreting, the abundance of data that these next-generation observational facilities will provide. To this end, this paper presents work in which we use numerical simulations to explore kinetic AGN jet feedback in the context of isolated gas-rich major mergers. The properties of SMBH host galaxies have been selected such that they are good analogues of galaxies at $z \approx 2$, where galaxies are expected to be highly gas-rich \citep{2020ForsterSchreiber+}, facilitating rapid AGN accretion and leading to elevated star formation rates. This likely results in powerful AGN feedback episodes and vigorous starbursts, meaning that outflows are expected to be prominent in such systems. We apply our spin-driven AGN jet feedback model \citep[see][]{2021Talbot+, 2022Talbot+}, to the black holes at the centres of both of the merging galaxies. Our simulations then follow the progression of the two galaxies and their AGN through the first passage and up to the final coalescence of their host galaxies. Using these simulations we explore the impact of AGN jet feedback on the stellar component and the gaseous haloes of the galaxies. Additionally, we explore how the AGN jets self-regulate when subject to the extreme environments present in such mergers.

The structure of the paper is as follows. In Section~\ref{Sec: BH model} we provide a brief overview of our black hole accretion and feedback model before explaining the changes made to the model that is used in this work, relative to that which was used in the simulations presented in \citet{2021Talbot+} and \citet{2022Talbot+}. Then, in Section~\ref{Sec: Additional physics}, we describe some of the additional physical processes modelled in these simulations. In Section~\ref{Sec: Set up} we describe how the different components of the galaxies are modelled and how we set up the initial conditions for the merger simulations. Our results are presented in Section~\ref{Sec: Results} and in Section~\ref{Sec: Discussion} we compare our results to those of previous works and discuss the limitations of our simulations. Finally, in Section~\ref{Sec: Conclusions}, we end with our conclusions.

%%%%%%%%%%%%%%%%%%%%%%%%%%%%%%%%%%%%%%%%%%%%%%%%%%%%%%%%%%%%%%%%%%%%%%%%%%%%%%%%%%%%%%

\section{Black hole accretion and feedback}
\label{Sec: BH model}
In this paper we use numerical simulations to explore the merger of two galaxies hosting active black holes. We adopt a novel sub-grid model for black hole accretion and feedback, as presented in \citet{2021Talbot+, 2022Talbot+}. This model couples an $\alpha$-disc accretion prescription \citep{2018Fiacconi+} to a feedback prescription for high resolution kinetic jets \citep{2017BourneSijacki} via the Blandford-Znajek mechanism. This model has been implemented into the moving-mesh code \textsc{arepo} \citep{2010Springel, 2016Pakmor+, 2020Weinberger+}.

We model black holes as sink particles and assume that they are surrounded by a sub-grid, thin, (potentially warped) $\alpha$-disc \citep{1973ShakuraSunyaev} which modulates the flux of mass and angular momentum across the black hole horizon. The mass and angular momenta of the black hole and $\alpha$-disc evolve due to inflows of gas from the surroundings, mutual Bardeen-Petterson torquing \citep{1975BardeenPetterson}, the launching of the Blandford-Znajek jet itself and the accretion of material by the black hole from the inner edge of the $\alpha$-disc. Accounting for these processes allows us accurately follow the mass and spin evolution of the black hole and self-consistently calculate the power and direction of the Blandford-Znajek jet \citep{1977BlandfordZnajek} that it launches. 

The power of a Blandford-Znajek jet depends on the magnetic flux that threads the black hole horizon. Due to the fact that our simulations do not include magneto-hydrodynamic effects and, more importantly, we are unable to resolve the black hole horizon, we instead make use of the predictions from general relativistic magneto-hydrodynamic (GRMHD) simulations and analytic arguments to fully parameterise the model \citep{2012Tchekhovskoy}.

The fluxes of mass and angular momentum onto the $\alpha$-disc are estimated from the properties of inflowing gas cells in the vicinity of the black hole. The jet is then launched by injecting mass,
energy and momentum into gas in the `jet-cylinder', centred on the black hole, with axis determined by the black hole spin direction. As stated above, the jet is mass-loaded from the inner edge of the $\alpha$-disc and, throughout this work, we assume that this occurs with a mass loading factor of unity. It should be highlighted, however, that injecting the jet energy, mass and momentum into the jet cylinders which already have a non-zero mass, means that the effective mass loading at the base of the jets in our simulation is primarily driven by the mass resolution of the jet cylinders.

\subsection{Gas circularisation condition for accretion}
\label{Subsec: Circ cond}
In our original implementation of the $\alpha$-disc model, inflowing gas could only flow onto the $\alpha$-disc if it was able to circularise and settle within the disc. We imposed this condition by allowing gas to accrete when the sector-averaged\footnote{The black hole accretion rate is calculated using the properties of the gas within a sphere with radius equal to the smoothing length of the black hole. To better model non-axisymmetric feeding of the black hole via coherent streams of material, we split this sphere into four sectors along the direction parallel to the angular momentum vector of the $\alpha$-disc and calculate the inflow rate separately within each sector. For further information, see section~$3.2$ of \citet{2021Talbot+}.} specific angular momentum of the gas was smaller than that of the outer edge of the disc.  

The imposition of such a condition requires that, in the simulation, the $\alpha$-disc is consistently being resolved by at least one cell. This was the case in the simulations presented in \citet{2021Talbot+} and \citet{2022Talbot+}, however, the simulations performed in this work have lower resolution (due to the significantly larger physical scales considered) and this condition is no longer satisfied. We, therefore, introduce a new circularisation condition for accretion which takes the lower resolution of our simulations into account.

Specifically, for each black hole in the simulation, at every timestep we calculate the smallest resolvable scale relevant to black hole accretion in each of the accretion sectors. When the jet is inactive, we take this to be the characteristic inflow radius in the relevant sector
\begin{equation}
\label{eq: rin}
   r_{{\rm in}, s} = \frac{\sum_{j=1}^{N_s}(3V_j/4\pi)^{1/3}\,W(d_j)}{\sum_{j=1}^{N_s}\,W(d_j)}\, ,
\end{equation}
where the sum is over the $N_s$ inflowing cells in the relevant sector, $V_j$ is the volume of the gas cell and $W(d_j)$ is a cubic spline kernel with compact support over the black hole smoothing length, $h_{\rm BH}$ \citep[for further information, see section~$3.2$ of][]{2021Talbot+}.

When the jet is active, however, the act of injecting mass, energy and momentum into the gas cells surrounding the black hole means that the hydrodynamics on scales smaller than this injection region are unresolved. We, therefore, choose the smallest resolvable scale to be the larger out of the radius of the jet cylinder and the characteristic inflow radius, given in Equation~(\ref{eq: rin}) above.

We then calculate the specific angular momentum of the inflowing gas in each sector. If it is less than that of the outer edge of the disc then the gas is allowed to flow onto the $\alpha$-disc, as per our original prescription. If, however, the specific angular momentum in the sector corresponds to a circularisation radius that is larger than the disc radius but smaller than this `smallest resolvable scale', we assume that the gas will be able flow onto the $\alpha$-disc but that, due to unresolved physical processes, its specific angular momentum is reduced to that of the outer edge of the $\alpha$-disc.

In doing so, we have implicitly assumed that all of this gas will reach the $\alpha$-disc. This may not necessarily be the case, but the efficiency by which unresolved processes transfer mass and angular momentum is not well understood and so we choose to assume `maximal efficiency' and highlight the fact that our inflow rate estimate may na{\"i}vely be taken to represent an upper limit to the accretion rate we would find, were we able to resolve the $\alpha$-disc. However, as the actual mechanism by which angular momentum is transferred in discs that are self-gravitating and star-forming gas is poorly understood, especially in the presence of supernova (SN) and AGN feedback, in reality our estimated mass flux on the $\alpha$-disc may be too low. Particularly as our simulations do not fully resolve the ISM structure and thus, we might expect the gas to be more clumpy and to reach much higher densities \citep[see e.g.][and Section~\ref{Subsec: eEOS} of this paper]{2003SpringelHernquist}.

\subsection{Targeted refinement criteria}
\label{Subsec: Ref}
In section~$3.7$ of \citet{2021Talbot+}, we detail several additional targeted refinement criteria\footnote{These criteria are applied in addition to the standard refinement criterion in \textsc{arepo} which ensures that the mass of gas cells remains approximately constant.} which we use to ensure the accretion flow as well as the jet injection and lobe inflation are sufficiently well resolved. In the work presented here, we make use of all of these refinement schemes, some of which we make improvements to, which we now briefly discuss.

We adopt a `black hole refinement scheme' which ensures that the spatial resolution within some chosen `refinement radius' around the black hole increases linearly towards the black hole \citep{2016CurtisSijacki}. We also use the jet cylinder refinement scheme \citep{2017BourneSijacki,2021Talbot+} as well as an additional refinement scheme which increases the spatial resolution in gas cells that contain a sufficiently high fraction (greater than $10^{-5}$) of a passive tracer that is injected with the jet. At later times in the simulations presented in this work, however, a significant volume of the simulation domain can be filled with jet material. In such scenarios, we found that using this jet lobe refinement scheme without modification ultimately leads to the formation of a numerically intractable number of cells. To ensure that we are able to follow the merger of these galaxies to completion, we instead base the jet lobe volume refinement on a decaying jet tracer with a fixed `decay' timescale of $100$~Myr. 

Additionally, the original refinement schemes did not prohibit the mass of a cell from becoming arbitrarily small. While this was not a problem for the setups considered in \citet{2021Talbot+} and \citet{2022Talbot+}, in the simulations presented here we found, however, that very low mass cells could lead to prohibitively small timesteps. These small cells were typically those into which jet energy had recently been injected and were often located close to a black hole. We therefore impose a mass floor, which acts on all cells in the black hole refinement region and all those subject to the jet volume refinement scheme (i.e. those with a decaying jet tracer fraction greater than $10^{-5}$). In the simulations presented in this work, we choose a mass floor of $10^3 \, {\rm M_\odot}$ which is approximately $100$ times smaller than the target gas mass of these simulations (see Section~\ref{Subsec: Initial conditions}).

\subsection{Energy injection in extreme environments}
\label{Subsec: Energy inj}
The radius of the cylinder into which we inject the jet feedback is calculated at each timestep and for each black hole, such that the mass contained within the cylinder is roughly constant and the cylinder is contained within the black hole smoothing length. In addition to this, we impose the condition that each half of the jet cylinder must be populated by at least $10$ gas cells. The jet cylinder refinement is usually able to ensure this condition is satisfied and the jet cylinder largely stays at constant mass.

Occasionally, however, these refinement schemes are not able to respond fast enough and, in order for this minimum cell number criterion to be satisfied, the jet cylinder would have to be too large. Such an occurrence is usually associated with a particularly powerful jet outburst. When this happens, instead of injecting the jet into too few cells, we store the mass, energy and momentum that should have been injected in this timestep, and inject in the next timestep when the cell number criterion is satisfied. Having too few cells in the jet cylinder is relatively rare in these simulations and and when it does happen the jet energy, mass and momentum usually only need to be stored for, at most, a few timesteps. 

%%%%%%%%%%%%%%%%%%%%%%%%%%%%%%%%%%%%%%%%%%%%%%%%%%%%%%%%%%%%%%%%%%%%%%%%%%%%%%%%%%%%%%

\section{Additional physics in the simulations}
\label{Sec: Additional physics}
In addition to the growth of, and feedback from SMBHs, the simulations presented in this work include models for several other key astrophysical processes, including star formation and evolution, chemical enrichment, the galactic winds driven by stellar feedback and primordial and metal-line gas cooling. In this section we provide a brief description of these processes and how they are modelled in the simulations. 

\subsection{Star formation, stellar feedback and galactic winds}
\label{Subsec: eEOS}
The simulations presented in this work do not have sufficient resolution to be able to capture small-scale physical processes, such as molecular cloud formation, their collapse and supersonic turbulence that are expected to lead to the development of a self-regulated ISM and the formation of stars. To model star formation and the pressurisation of the ISM due to unresolved SNe we, therefore, describe the star-forming gas using an effective equation of state (eEOS), largely following the prescription described in \citet{2003SpringelHernquist} (see also \citet{2013Vogelsberger+,2018Pillepich+}). To avoid overpressurising the ISM gas, which would lead to unrealistically `smooth' ISM, we interpolate between the eEOS detailed in \citet{2003SpringelHernquist} and an isothermal equation of state with temperature $10^4$~K, using an interpolation parameter $q_{\rm EOS} = 0.5$. We assume a Chabrier initial mass function (IMF) \citep{2003Chabrier} and model mass and metal return from Type Ia and Type II SNe, and asymptotic giant branch (AGB) winds \citep[for further details see][]{2013Vogelsberger+}.

In these simulations, stars are prevented from forming in the refinement regions surrounding the black holes. The additional refinement (as described in Section~\ref{Subsec: Ref}) means that gas cells in this region can have masses that are significantly smaller than cells in the majority of the simulation domain. Were stars to be allowed to form in this region, then the resulting wide range of star particle masses could lead to spurious N-body heating effects, which could potentially affect the temperature of the gas cells close to the black hole, as well as lead to the `ejection' of light stellar particles.

The simulations in this work use a threshold density for star formation of $\rho_{\rm th} = 0.03 \, {\rm cm^{-3}}$ and $t_{0}^{*} = 12.6$~Gyr, using the notation of \citet{2003SpringelHernquist}. Our gas consumption timescale is longer than the fiducial value presented in \citet{2003SpringelHernquist}, to better match that which is expected in gas-rich galaxies at $z\approx 2$. Additionally, we would expect these galaxies to have significantly higher levels of turbulence in their ISM than local galaxies. We chose to model this enhanced pressurisation by using a lower value of $\rho_{\rm th}$ than the value presented in \citet{2003SpringelHernquist}. We additionally use a sub-grid model for galactic winds that are launched directly from the star-forming phase, again, largely following the formalism outlined in \citet{2003SpringelHernquist} (see also \citet{2013Vogelsberger+,2018Pillepich+}). Our simulations use a wind mass loading factor of $\eta = \dot{M}_{\rm w}/\dot{M}_* = 1$ where $\dot{M}_{\rm w}$ is the mass flux into the wind and $\dot{M}_*$ is the star formation rate. The initial velocity of a wind cell is determined under the assumption that some fraction of the energy available from SNe (in this work, we assume a value of $10$ per cent) drives the winds and its direction is randomly oriented along $\boldsymbol{v}\times\nabla \phi$ (i.e. in a bipolar manner) where $\boldsymbol{v}$ is the velocity of the gas cell and $\phi$ is the local gravitational potential \citep{2003SpringelHernquist}. Note that this differs from the wind launching model used in the Illustris simulations \citep{2013Vogelsberger+} and from that used in the subsequent Illustris TNG simulations \citep{2018Pillepich+}.

\subsection{Radiative cooling and heating}
\label{Subsec: Rad cool heat}
In these simulations, primordial species undergo radiative cooling according to the primordial atomic network described in \citet{1996Katz+}. Metal-line cooling is implemented in the simulations using pre-calculated look-up tables, generated by the photoionisation code \textsc{cloudy} \citep{2017Ferland+}. These tables provide cooling rates down to temperatures of $10$~K and are calculated for a solar composition gas and then scaled linearly with the metallicity of the cell\footnote{Throughout this work, we assume $Z_\odot = 0.0127$.}. The ultraviolet background (UVB) is modelled as a time-dependent, spatially uniform radiation field which injects heat into the gas. These simulations use the UVB intensity from \citet{2009Faucher-Giguere+} and include the self-shielding correction from \citet{2013Rahmati+}. Since the simulations presented in this work are intended to model galaxy mergers at cosmic noon, we ensure that the time-dependent cooling processes correspond to $z=2$ values. 

When using the eEOS model, described in Section~\ref{Subsec: eEOS} above, typically metal-line cooling is usually only considered in gas above $10^4$~K \citep{2013Vogelsberger+}. There is no reason, however, why the gas at densities lower than the star formation threshold should not be able to cool down to these temperatures. We, therefore, considered two different cooling prescriptions: In the `fiducial' case, metal-line cooling is only effective down to $10^4$~K. In the `additional cooling' case, metal-line cooling is effective down to $10^4$~K for gas with density higher than the star formation threshold, while gas at lower densities is allowed to cool down to $10$~K via this channel. 

%%%%%%%%%%%%%%%%%%%%%%%%%%%%%%%%%%%%%%%%%%%%%%%%%%%%%%%%%%%%%%%%%%%%%%%%%%%%%%%%%%%%%%

\section{Simulation set-up}
\label{Sec: Set up}
In this work we use simulations to explore the major merger of two gas-rich galaxies hosting AGN-driven jets. We consider two simulations which, throughout this work, we refer to as the `jet' and `no-jet' simulations. The black holes in the `jet' simulation launch jets via our spin-driven jet model, whereas the black holes in the `no-jet' simulation accrete via sub-grid $\alpha$-discs but do not launch jets. Both of these simulations use the `additional cooling' prescription, as described in Section~\ref{Subsec: Rad cool heat}, wherein gas with density lower than the eEOS threshold can cool down to $10$~K via metal-line cooling. This low temperature cooling acts to somewhat increase the mass in cool and star-forming phases, particularly in the gas that is cooling out of the outflow. This facilitates somewhat higher black hole inflow rates and, ultimately, higher jet powers. On the other hand, the more powerful jets more readily heat and destroy the cool gas that forms. Hence, overall we find very modest qualitative difference with respect to the `fiducial' cooling simulation, and we only provide analysis of the data from the `additional cooling' simulations in this work.

In this section, we detail the properties of the merging galaxies and explain how the initial conditions for the simulations were set up.

\subsection{Initial conditions}
\label{Subsec: Initial conditions}
To create the initial conditions for the merger, we first set up a single isolated system of mass $M_{200} = 10^{12} \, {\rm M_\odot}$ using the procedure outlined in \citet{2005Springel+}. The system consists of a dark matter halo, a disc of gas and stars, a stellar bulge, a hot, gaseous halo and a black hole.

Specifically, the stellar and gaseous disc have total masses of $1.64\times10^{10} \,  {\rm M_\odot}$ and $2.46\times10^{10} \,  {\rm M_\odot}$, respectively, giving a gas fraction of approximately $0.6$. Both discs are modelled as having exponential surface density profile with a scale length of $2.6 \,  {\rm kpc}$. The stellar disc has vertical structure corresponding to that of an isothermal sheet with scale-height of $0.26 \,  {\rm kpc}$, while the gas disc is initialised with a temperature of $10^4$~K and its vertical structure is set up to ensure hydrostatic equilibrium. The dark matter halo structure and that of the hot halo are modelled using a Hernquist profile \citep{1990Hernquist} whose structure closely follows that of an NFW profile \citep{1997Navarro+} with concentration parameter, $c = 9$ and a spin parameter, $\lambda = 0.033$. The stellar bulge is assumed to be spherical and follow a Hernquist profile \citep{1990Hernquist}, with $a = 0.26 \,  {\rm kpc}$ and with a total mass of $8\times10^9 \,  {\rm M_\odot}$. At the centre, we place a black hole of mass $10^7 \, {\rm M_\odot}$, which is consistent with the black hole mass-bulge mass relation.

The dark matter halo is sampled by collisionless particles of mass $3\times 10^6 \, {\rm M_\odot}$, the star particles in the disc and bulge are of mass $8.2\times 10^4 \, {\rm M_\odot}$ and  $1.6\times 10^5 \, {\rm M_\odot}$, respectively, and the initial mass of the gas cells in the halo and disc is $1.23\times 10^5 \, {\rm M_\odot}$. The dark matter particles have gravitational softening lengths of $2$~kpc while that of stars in the disc and bulge is $60$~pc. The black hole has a softening length of $150$~pc and that of the gas is treated adaptively, but with a minimum of $60$~pc.

In these simulations we do not allow the black hole particles to merge and, instead, they form a binary. We do so as the standard merger prescription in \textsc{arepo} (which instantaneously merges black holes within their own smoothing lengths) is too efficient and does not account for realistic stellar and gaseous hardening processes which will likely cause the black holes to merge on a significantly longer timescale than this `instantaneous merger' prescription. We wish to highlight, however, that since the gravitational force law used in the simulations is softened, it will not be possible to accurately follow the hardening of the binary to separations smaller than twice the black hole softening length ($\sim 300$~pc). We did, however, perform analogous simulations with smaller black hole softening lengths and confirmed that the results presented in this paper are robust to such changes. In these simulations, the dynamics of the black holes evolve according to the gravity solver and we do not use `repositioning' techniques. This is reasonable as we are in the regime where relevant dynamical friction processes are well resolved.

It should also be pointed out that we do not pre-enrich the gas disc, nor the hot halo of the galaxies since significant star formation and subsequent metal return rapidly enriches the disc. The galactic winds and jets then transport this enriched gas into the halo, such that the metallicity of the circumgalactic medium (CGM) is consistent with \citet{2015Suresh+} within a few Myr.

Having set up initial conditions for an isolated galaxy, as described above, we then create initial conditions for the merger by combining two such galaxies. The initial positions and velocities of the galaxies are chosen such that they are on a prograde, parabolic orbit with the plane of both discs lying in the orbital plane. The initial displacement of the galaxies is set to $2\times R_{200} = 325$~kpc and the expected pericentric distance is set to $2$~kpc such that significant funnelling of gas towards the centres of the galaxies is expected during first passage, as a result of the powerful tidal torques in this configuration \citep[see e.g.][]{2008Lotz+}. This merger setup is then placed in a cubic simulation domain with a side-length of $1$~Mpc. 

Since the gas discs are not guaranteed to be in equilibrium and since we wish to focus on the behaviour of the system as the galaxies approach first passage and beyond, we run an initial simulation of $\sim1$~Gyr and then use the output of the simulation as initial conditions for relevant simulation with the $\alpha$-discs and jets. At the end of this simulation the separation between the galaxies is $\sim 25$~kpc and the discs are already showing signs of significant tidal deformation.

In this initial simulation we do not use our black hole accretion and jet model. During the time in which the galaxies are coming into equilibrium, the properties of the jets that would form would likely be driven by our choice of initial conditions which could potentially bias our results (for example if the jets were to significantly perturb the galactic discs). Instead, during this time, we allow the black holes to accrete at a fraction of their Bondi-Hoyle-Lyttleton rate, where the fraction is chosen to ensure that the final black hole masses remain consistent with the black hole mass-bulge mass relation. The output of this simulation is then used as initial conditions for the `production' simulations which use the full black hole spin-driven jet model (or just the $\alpha$-disc model for the simulation without jets). The rest of this paper will focus on the `production' simulations and will not discuss these `initial' simulations any further.

In the `production' simulations, we choose the initial spins of both black holes to have a magnitude of $0.7$ and direction parallel to the $z$-axis of the simulation domain. As the simulations progress, these quantities are then free to evolve according to our model. The initial masses of the $\alpha$-discs are chosen to be $10^5 \, {\rm M_\odot}$ and their angular momenta are also initialised parallel to the $z$-axis. 

Since our choice of initial $\alpha$-disc mass does not guarantee that the disc will be in equilibrium with the accretion flow, we wait $10$~Myr at the start of the simulation before allowing jets to be launched\footnote{Note that after these $10$~Myr have elapsed, we reset the spin direction and magnitude of the black holes, as well as the angular momentum direction of the $\alpha$-discs so that the (not necessarily realistic) evolution of these quantities as the $\alpha$-disc masses come into equilibrium does not lead to the launching of jets with unphysical properties.}.

%%%%%%%%%%%%%%%%%%%%%%%%%%%%%%%%%%%%%%%%%%%%%%%%%%%%%%%%%%%%%%%%%%%%%%%%%%%%%%%%%%%%%%

\section{Results}
\label{Sec: Results}

\subsection{Qualitative overview of the simulations}
\label{Subsec: Qualitative}

\begin{figure*}
    \centering
    \includegraphics[width=0.98\textwidth]{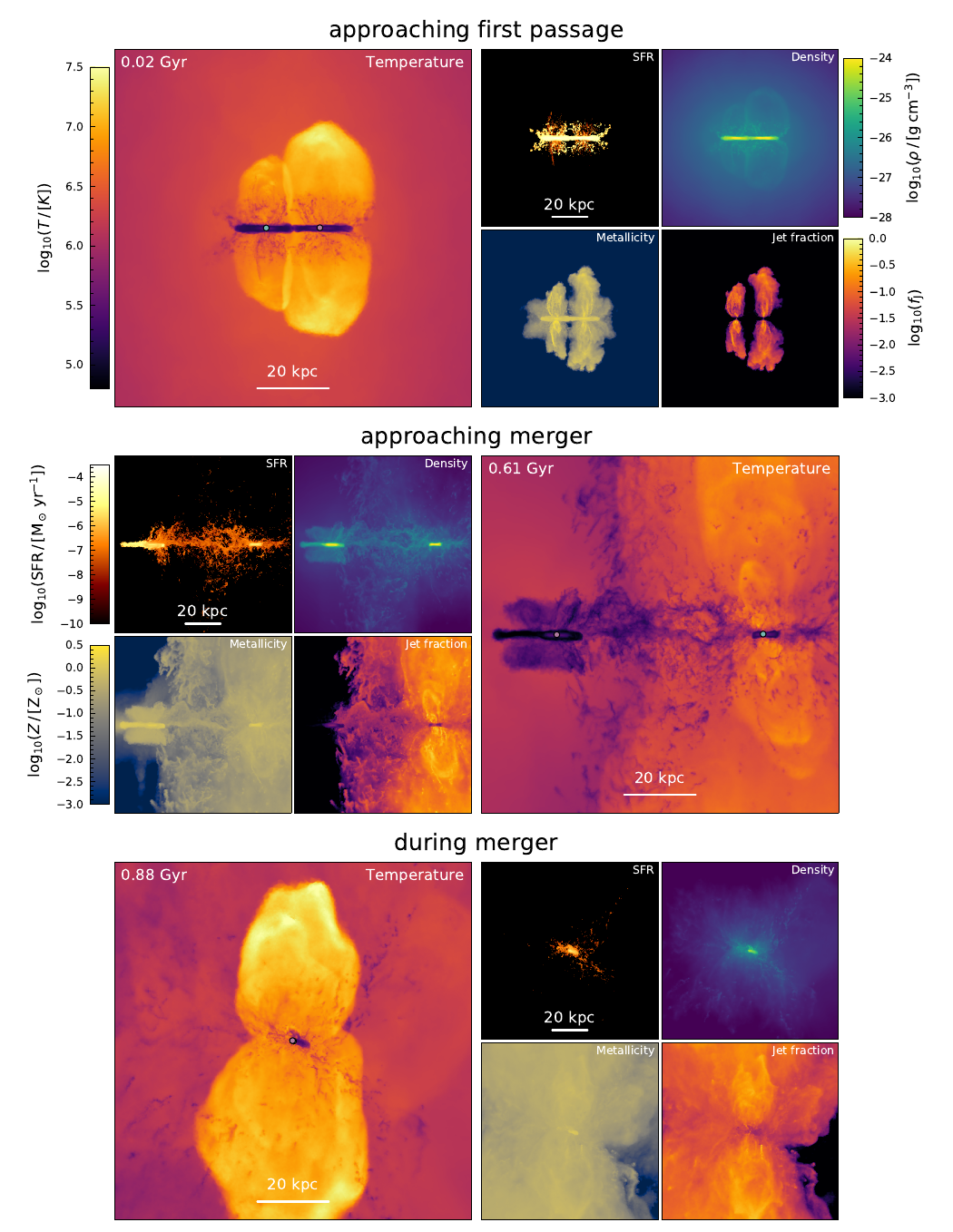}
    \vspace{-0.3cm}
    \caption[Projections that give a qualitative overview of the merger simulation with jets]
    {A qualitative overview of the simulation with jets. In each of the three rows, the main panel shows a projection of the gas temperature. The four smaller panels show (clockwise, starting in the top left) the star formation rate, the gas density, a passive tracer injected with the jet and the gas metallicity. Each row shows the state of the system at an important moment in the merger process (as indicated by the row headings) and the time to which this corresponds is shown in the top left hand corner of each of the main panels. The position of the black holes are indicated by coloured markers in the temperature panel, using the same colours as those of the lines in Fig.~\ref{fig: subgrid multiple}. Note that in the bottom row, the black hole separation is very small so the markers are largely overlapping.}
    \label{fig: Various quantites}
\end{figure*}

We begin our analysis by giving a qualitative overview of how the galaxy merger progresses when jets are present. We do so with the aid of Fig.~\ref{fig: Various quantites} which shows projections of various gas properties for the simulation with jets. Each of the three rows shows the state of the system at a key moment in the merger process (as indicated by the row headings) and the corresponding time is indicated in the top left-hand corner of each of the main panels. Each projection is centred on the midpoint of the separation vector of the black holes and has axis that is perpendicular to the plane of the galaxies\footnote{For simplicity, throughout the rest of this work, we will refer to this configuration as `edge-on', and we will refer to projections that are centred on the centre of the simulation domain and have axis that is parallel to the $z$-axis of the simulation as `face-on'.}.

In the first row of Fig.~\ref{fig: Various quantites}, the two galaxies are approaching first passage. Hot, diffuse cocoons are clearly visible in the temperature map, inflated by the young jets which have velocities of order $10^{4} \; {\rm km \, s^{-1}}$ at launch. Additionally, there is a region of enhanced temperature where the cocoons overlap and shock. Due to slight differences in the masses of the black holes and their accretion flows, the initial powers of the jets are not the same (see Section~\ref{Subsec: Jet properties}) and one of the jets is slightly more powerful at first. This results in a slightly larger cocoon driven by this jet and hotter hotspots.

In addition to the hot gas in the cocoons, there is also gas with temperatures of $10^4-10^5$~K associated with the galactic wind outflows. Inspection of the SFR projection shows that star forming gas is present throughout much of the galactic discs and, additionally, in some of the dense outflowing material associated with the galactic winds. From the metallicity and jet tracer maps, it is also clear that the jet material is highly metal enriched, which highlights the fact that jets can act to expel material that has been enriched by stellar evolution from the galactic discs and draw it up in the outflows.

In the second row of Fig.~\ref{fig: Various quantites}, more than half a Gyr has elapsed and the galaxies are now approaching their final coalescence. The strong tidal forces experienced by the galaxies after their first passage have led to the formation of extended tidal tails and a bridge of gas (and stars) between the two galaxies. Now the other jet is considerably more active (this will be quantified in Section~\ref{Subsec: Jet properties}). This more active jet is heating much of the gas in its vicinity and is significantly metal-enriching the halo, whilst the largely `quiescent' jet is currently having very little effect on the gas beyond its immediate vicinity. 

There is also significant amount of warm and cold gas that can be seen falling back towards the midplane, in the region between the two galaxies. This gas is associated with earlier jet activity where the jet-driven outflows have decelerated and cooled, and begun to fall back into the potential well of the system and it is this infall of cooling gas that is actually responsible for feeding the black hole and causing the jet power to increase. At this time, star formation is still occurring throughout the majority of the galactic discs and in the stellar winds. It is also clear that star forming gas is still present in the dense outflows as well as the inflows associated with the jet activity. 

The fact that we find star forming gas out of the plane of the galactic discs highlights the possibility that stars may exist in these jet-driven outflows. Indeed, we do also find that stars are present and form in the outflows and inflows associated with jet activity in our simulations. The star formation rate in gas associated with the outflow\footnote{i.e. in the region $20-100$~kpc above and below the midplane and within a cylindrical radius of $100$~kpc from the centre of the box (see Section~\ref{Subsubsec: Outflow properties}).} can reach $\sim 10^{-3} \; {\rm M_\odot \, yr^{-1}}$ although outbursts from the jet can reduce this significantly. Additionally, the mass of stars in the outflow can reach $\sim 10^7 \;{\rm M_\odot}$ by the end of the simulation. While non-negligible, the mass of outflowing stars (and star-forming gas) is significantly smaller than the stellar mass of the disc (see Section~\ref{Subsec: Stars}). We do wish to emphasise, however, that the model we use for star formation and stellar feedback (see Section~\ref{Subsec: eEOS}) is not necessarily applicable to the formation of stars in galaxy-scale outflows. The presence of stars and star forming gas in the outflows in our simulations should, therefore, be interpreted as indicative of the \textit{possibility} of finding stars in these outflows. To quantify this further, a more accurate model for star formation and stellar feedback would be required.

In the final row of Fig.~\ref{fig: Various quantites}, the galaxies have reached coalescence and have formed a dense compact system that can be identified in the temperature, density and SFR projection and is surrounded by significantly metal enriched gas. The merger of the galaxies has triggered a powerful jet outburst extending up to $50$~kpc away from the merger remnant. One other interesting feature to highlight is that the axis of the jet-driven bipolar outflow is no longer aligned with the vertical, as is clearly visible in the temperature map. This change in the orientation of the large-scale outflow arises due to the fact that the spin of the black hole associated with the most active jet has been torqued and is now somewhat inclined to the vertical as, during the course of the simulation, this black hole is fed by gas with misaligned angular momenta. This will be further discussed and quantified in Section~\ref{Subsec: Jet properties}. 

\begin{figure}
    \centering
    \includegraphics[width=\columnwidth]{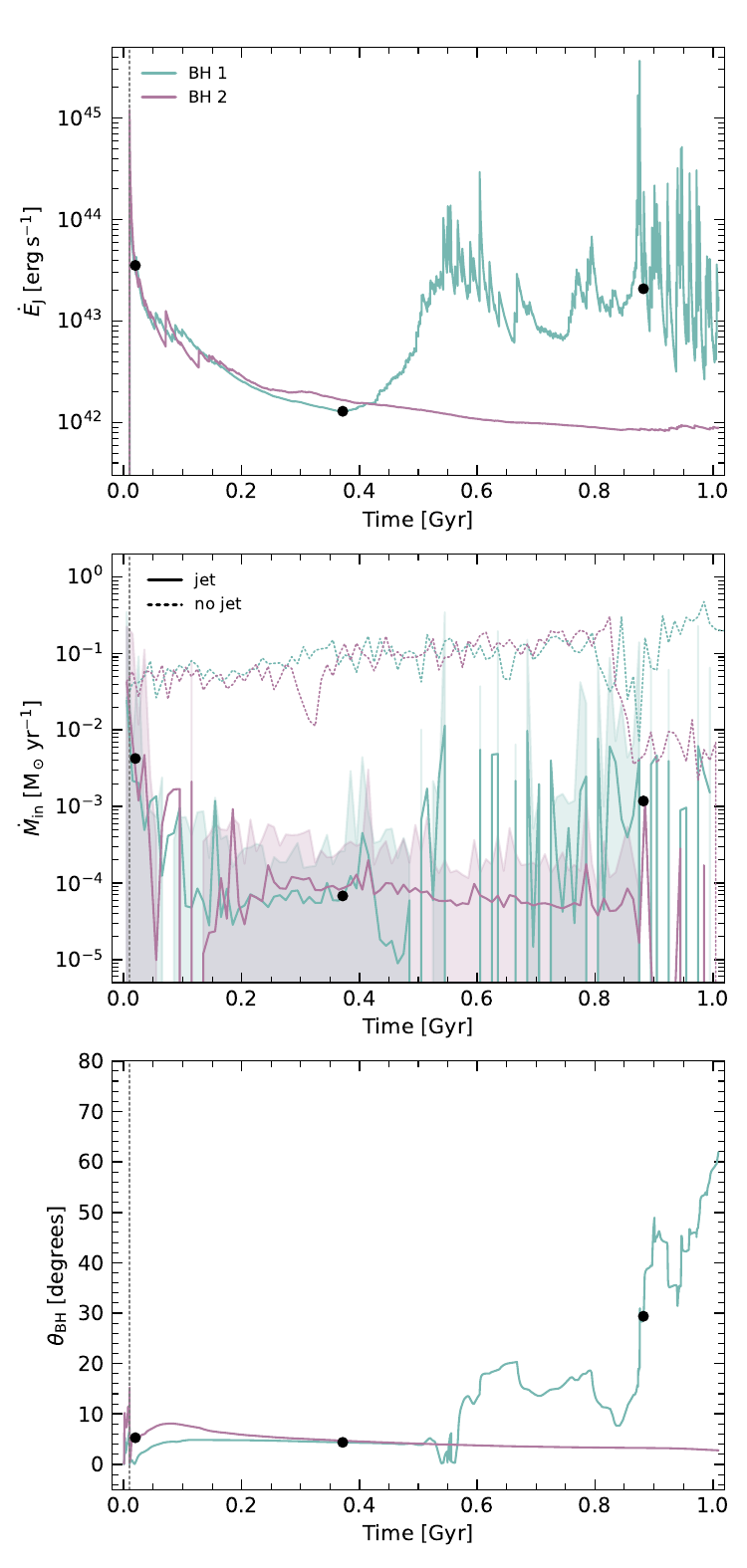}
    \caption[Time evolution of the accretion flow and black hole and jet properties]
    {\small The evolution of the jet powers (top panel), the time-averaged mass inflow rates onto the sub-grid $\alpha$-discs (middle panel) and the evolution of the inclinations of the jets (i.e. black hole spins) to the vertical (bottom panel). In the middle panel, the shaded regions bound the range of observed inflow rates and the dotted lines indicates the inflow rates measured in the analogous simulation without jets. The properties of the two black holes in each simulation are indicated by the different colour curves (see the legend in the top panel). These colours match those of the markers in Fig.~\ref{fig: sam and temp} that indicate the location of each black hole, and the black markers indicate the properties of the relevant black hole at each of the times shown in Fig.~\ref{fig: sam and temp}. The grey, vertical, dotted line indicates the time at which the jets first become active.}
    \label{fig: subgrid multiple}
\end{figure}

\subsection{Black hole and jet properties}
\label{Subsec: Jet properties}

Having qualitatively explored the key processes that influence the evolution of the merger in the presence of jet-driven outflows, we now turn to more quantitative analyses, first examining the properties of the black holes and jets themselves. 

Fig.~\ref{fig: subgrid multiple} shows the evolution of the jet powers, the mass inflow rates onto the sub-grid $\alpha$-discs, averaged in $10$~Myr windows, and the inclinations of the jets (i.e. black hole spins) to the vertical. The initial jet outburst is more powerful for BH~$2$ than BH~$1$. This is consistent with the discussion in Section~\ref{Subsec: Qualitative} where we highlighted the fact that, in the first row of Fig.~\ref{fig: Various quantites}, the cocoon associated with the jet on the right is slightly larger. For the rest of the simulation, however, BH~$1$ is the most active: in the period of time before the galaxies coalesce, the power of the jet associated with BH~$2$ is maintained at a few times $10^{42}\, {\rm erg\,s^{-1}}$ while that of the jet associated with BH~$1$ is able to reach powers as high as $\sim10^{45}\, {\rm erg\,s^{-1}}$ during the final coalescence\footnote{Note that the jet power is modulated somewhat by the sub-grid accretion disc, as opposed to being directly linked to the inflow rate.}, resulting in the highly energetic outflow seen in the bottom row of Fig.~\ref{fig: Various quantites}. 

To understand the general behaviour of these jet powers, consider the middle panel of Fig.~\ref{fig: subgrid multiple}, which shows the time-averaged mass inflow rates onto the sub-grid $\alpha$-discs. For the vast majority of the time before the final coalescence of the galaxies (which begins at $\sim 0.8$~Gyr), gas is able to flow onto the sub-grid $\alpha$-disc of both black holes. From the shaded region around each line, which bounds the range of mass inflow rates, we can see that, while gas is being fairly consistently supplied to the $\alpha$-disc, there are short periods of time without inflow. BH~$2$ experiences fairly steady inflows that are able to maintain a reasonably constant jet power, while BH~$1$ experiences much more variable inflow rates that result in larger fluctuations in the jet power. These bursts of inflow are driven by the cool gas that is raining back onto the galaxy, as discussed in Section~\ref{Subsec: Qualitative}. The fact that there tends to be one black hole that is accreting more, thus launching a more powerful jet initially comes about due to the lack of perfect symmetry of gas properties around the two black holes. Additionally, the launching of a powerful jet leads to further accretion by inducing the `rain' of gas which cools out of the hot jet-driven outflow and can then further enhance the jet power, leading to a situation reminiscent of a positive feedback loop. Throughout the final coalescence of the galaxies, bursty inflows are able to persist, primarily for BH~$1$, as is reflected in its high jet power. 

The behaviour of the inflow rate during the coalescence of galaxies is markedly different from what would be found when using accretion rate estimates such as Bondi-Hoyle-based prescriptions, which would be approximately $1$-$2$ orders of magnitude higher. Our requirements that the gas be flowing towards the black hole and have angular momentum such that it is able to circularise and settle on the $\alpha$-disc means that inflow is not always guaranteed to occur (even in simulations without jets during the final coalescence, where the inflow rates become sporadic), despite the fact that gas is always present in the vicinity of the black holes\footnote{It should be stressed, however, that even when there is no inflow onto the sub-grid $\alpha$-disc, the black holes in all of the simulations are always accreting as they never completely drain the gas from their $\alpha$-disc.}. These inflows lead to black hole growth rates that are typically $\sim 10^{-4} \, {\rm M_\odot \, yr^{-1}}$ but can reach $\sim 10^{-1} \, {\rm M_\odot \, yr^{-1}}$ when the inflow rates are highest. This rather modest black hole growth is in part due to the fact that the black hole spin energy is used to power the jets \citep[see e.g.][]{1977BlandfordZnajek, 2012Tchekhovskoy, 2022Talbot+} along with the effect that the jets have on the surrounding gas inflows. Indeed, in our simulations without jets, the inflow rates are $2$-$3$ orders of magnitude higher than those found in the simulations with jets (see the dashed lines in the middle panel of Fig.~\ref{fig: subgrid multiple}), reaching typically  $\sim 10^{-1} \, {\rm M_\odot \, yr^{-1}}$. This highlights the considerable impact that jets can have on the feeding of the black holes, which is discussed further in Section~\ref{Subsec: BH feeding}. It also clearly indicates that our accretion model (and the circularisation condition) onto the $\alpha$-disc allows for high and sustained accretion rates onto the black holes prior to the coalescence as expected in merging gas-rich galaxies.

One additional point worth highlighting is that the motions of the black holes at late times in these simulations will be affected by the numerical softening of the gravitational force, as discussed in Section~\ref{Subsec: Initial conditions}. To robustly measure the accretion rate onto the black holes after the formation of the binary would require a more accurate model for the black hole orbital dynamics. 

We now turn to the evolution of the jet direction (i.e. the black hole spin direction), which is shown in the bottom panel of Fig.~\ref{fig: subgrid multiple}. It is clear that the spin of BH~$2$ is not significantly torqued over the course of the simulation and remains approximately vertical throughout. The jet associated with BH~$1$, however, shows quite different behaviour. The direction of this jet is much more variable and, after $\sim1$~Gyr, it is inclined by $\sim 60^\circ$ (this is consistent with the misalignment of the outflow that can be seen in Fig.~\ref{fig: Various quantites} and was discussed in Section~\ref{Subsec: Qualitative}).

The significant torquing of the jet associated with BH~$1$ arises due to mechanism by which this black hole is fed. This black hole primarily accretes gas that has cooled out of the outflow and fallen back onto the galaxy and this material does not necessarily have angular momentum perpendicular to the plane of the galactic discs. As discussed in Section~\ref{Subsec: Qualitative}, comparatively less cold and cooling gas falls back onto the galaxy that hosts BH~$2$ and so, during the course of the simulation, it is largely being fed by gas that has circularised in the disc, causing little change to its angular momentum direction. Furthermore, during coalescence, the discs of the galaxies are disrupted as gas is violently shocked and subject to extreme tidal torques meaning that the gas available for accretion does not necessarily have angular momentum perpendicular to the orbital plane. This is likely responsible for the rapid changes to the jet direction during this time.

It is worth returning once more to the differing behaviours of the two black holes in this simulation. The enhanced level of activity of one of the two black holes arises due to the fact that it is able to enter a `feedback loop' whereby an initial powerful jet outburst induces the `rain' of gas which cools out of the hot jet-driven outflow and is then accreted by the black hole, thus further enhancing the jet power. Once this black hole enters this `feedback-loop', it remains in this state while there is gas still available. The fact that this behaviour is only seen for one of the two black holes highlights the fact that it only requires a small change to the inflow properties to allow a black hole to enter this regime.

\subsection{Black hole feeding and local gas properties}
\label{Subsec: BH feeding}

\begin{figure*}
    \centering
    \includegraphics[width=0.98\textwidth]{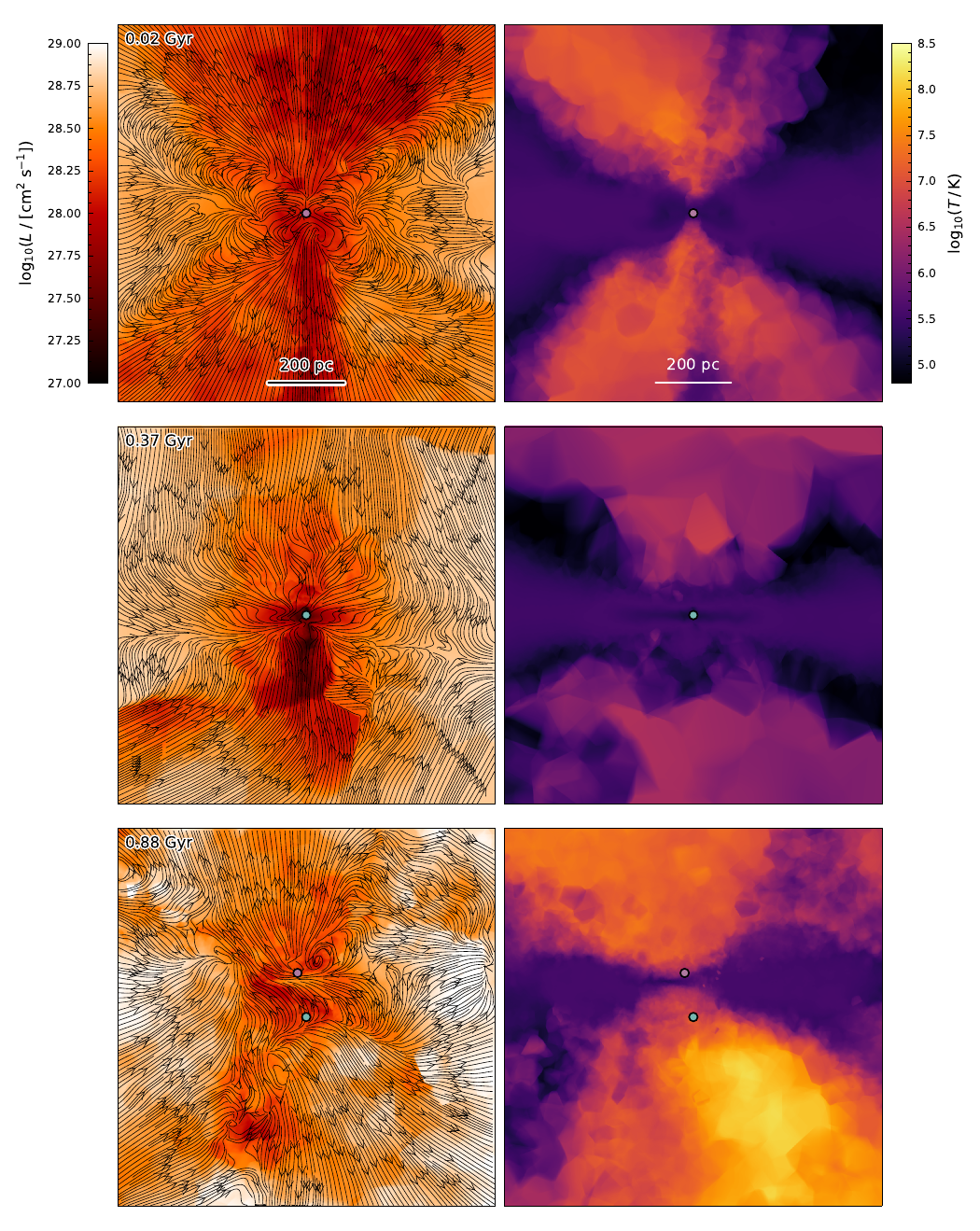}
    \vspace{-0.5cm}
    \caption[Projections of the specific angular momentum and temperature of the gas in the vicinity of the black holes]
    {Thin projections (of dimension $1\times 1\times 0.4$~kpc) for the simulation with jets which show, in the left-hand column, the magnitude of the specific angular momentum of the gas where the arrows correspond to the streamlines of the velocity field, and in the right-hand column, the temperature of the gas. In the first row, the projections are centred on BH~$2$ and then in the second and third rows, the projections are centred on BH~$1$. In each panel, the locations of the black hole(s) are marked with a coloured dot, using the same colours as those of the lines in Fig.~\ref{fig: subgrid multiple}.}
    \label{fig: sam and temp}
\end{figure*}

\begin{figure*}
    \centering
    \includegraphics[width=0.9\textwidth]{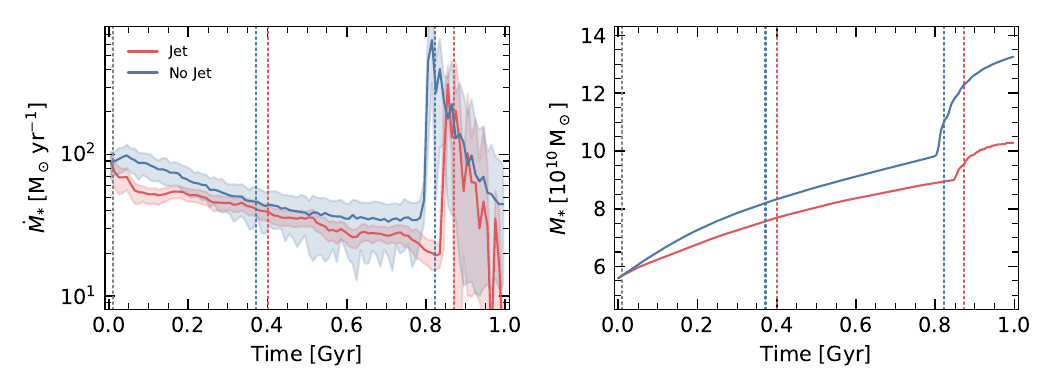}
    \caption[Time evolution of the star formation rate and the mass of newly formed stars]
    {\small Time evolution of the star formation rate (SFR) averaged in $10$~Myr windows (left panel) and the mass of newly formed stars (right panel). The red/blue curves correspond to simulations with/without jets. In the panel that shows the SFR, the shaded regions indicate the $10$-$90$ percentile range. Note that in the panel that shows the mass of newly formed stars, the curves correspond to the cumulative mass of stars that have ever formed (i.e. ignoring stellar evolution). In each panel, the set of red and blue vertical dotted lines at $\sim 0.4$~Gyr indicate the time at which the galaxies reach the apocentre of their orbit. The set of vertical lines at $0.8-0.9$~Gyr indicate the first time the SMBH separation is less than $2$ times the smoothing length of the black holes. The grey, vertical, dotted line indicates the time at which the jets first become active.}
    \label{fig: SFR}
\end{figure*}

In the previous section, we examined the behaviour of the gas inflow rate onto the black hole-$\alpha$-disc system. Here, we explore some of the physical processes responsible for modulating the accretion rates. We do so using Fig.~\ref{fig: sam and temp}, which shows thin projections ($1\times 1\times 0.4$~kpc) of the magnitude of the specific angular momentum of the gas, overlaid with arrows that show the streamlines of the velocity field (left column) and the temperature field (right column) for the jet simulation. In the first row, the projections are centred on BH~$2$ (the most powerful jet at this time, see Fig.~\ref{fig: subgrid multiple}) and then in the second and third rows, the projections are centred on BH~$1$. In each panel, the locations of the black hole(s) are marked with a coloured dot.  The first row shows the state of the gas during the initial outbursts of the jets. The second row corresponds to the time when gas is raining back down onto the galaxy and feeding the black hole, but the jet power has yet to increase significantly. Finally, the third row shows the state of the system during the coalescence of the galaxies and the formation of the black hole binary. In Fig.~\ref{fig: subgrid multiple}, the jet power, mass inflow rate and jet inclination at these times are indicated with a filled black dot, for the relevant black hole.

In the the top row, we can clearly identify the hot, jet-driven outflow, which consists of comparatively low specific angular momentum gas. The jet-driven outflow is propagating away from the galactic disc which, as is visible in the temperature map, has not yet been significantly disrupted. The streamlines of the gas in the disc indicate the presence of turbulence and, additionally, that a significant amount of disc gas is moving radially towards the black hole. At this time, the inflow rate onto the sub-grid $\alpha$-disc is still fairly high (see Fig.~\ref{fig: subgrid multiple}), which we can, therefore, infer as being largely due to these `secular' inflows of gas from the galactic disc. In the temperature map we can identify regions of cool gas that likely originated in the galactic disc within the hot outflow and inspection of the streamlines indicate that this gas has been entrained and is radially outflowing.  

In the middle row of Fig.~\ref{fig: sam and temp}, the projections are now centred on the other black hole (BH~$1$). The power of its jet is currently at a minimum (see Fig.~\ref{fig: subgrid multiple}) and this energy injection rate is not high enough to drive any significant outflow, but rather, the streamlines indicate that gas is flowing from the halo towards the galaxy and black hole. Interestingly, some of this inflowing halo gas has a comparatively low specific angular momentum and temperature. As discussed in Section~\ref{Subsec: Qualitative}, the jet-driven outflows act to seed thermal instabilities in the halo gas which, ultimately, results in cold and warm gas raining down on this galaxy. Having inspected specific angular momentum projections of the halo gas (not shown here), analogous to those shown in Fig.~\ref{fig: sam and temp}, we indeed find that this cool inflowing gas typically has low enough specific angular momentum such that it can circularise onto our sub-grid $\alpha$-disc and feed the black hole. This lends weight to our observation that, at this time, a non-negligible contribution to the gas that feeds the black hole comes from gas that condenses out of the halo and falls back down on the galaxy.

In the final row of Fig.~\ref{fig: sam and temp}, the galaxies have now reached coalescence. The projections are still centred on BH~$1$ (the green circle) but the other black hole is now also visible. We have already seen that BH~$1$ launches the more powerful jet during galaxy coalescence and is, thus, primarily responsible for driving the fast, hot outflow. The other jet is also active at this time, albeit with a jet power that is more than an order of magnitude lower (see Fig.~\ref{fig: subgrid multiple}). It is, therefore, the \textit{interaction} of these two jets that results in the complex velocity field that can be seen in the velocity streamlines at this time.

In \citet{2022Talbot+}, we performed similar analysis of the properties of gas feeding black holes in the presence of jets, focusing on isolated, non-interacting galaxies. We showed that that black holes can be fed via gas inflows from the surrounding (circumnuclear) disc but that when the jets are active, they drive backflows which can also play a crucial role in funneling low specific angular momentum gas towards the black hole \citep[see also][]{2010Antonuccio-DeloguSilk,2014Cielo+,2017BourneSijacki}. Here we confirm that both of these processes contribute to black hole feeding: the streamlines indicate that gas in the galactic disc is being funneled towards the black hole and the presences of vortices, associated with `small-scale' jet-driven backflows can also be seen in Fig.~\ref{fig: sam and temp}, drawing gas back towards the black holes. 

We find, however, that when radiative cooling processes are included in the simulations, jet-induced condensation of gas that then falls back onto the galaxy can also responsible for providing fuel for the black hole \citep[see also][]{2012McCourt+, 2013Gaspari+,2015Gaspari+,2015Prasad+,2019Wang+}. Additionally, whilst not shown in Fig.~\ref{fig: sam and temp}, when the galaxies pass each other for the first time during the final coalescence (at about $0.85$~Gyr), the extreme merger torques cause the gas in the galactic discs to lose angular momentum, thus forming compact, dense structures and this decrease in specific angular momentum, ultimately, leads to enhanced accretion rates, particularly in the case of BH~$1$. It is important to highlight that the interplay of all these processes is what, ultimately, determines black hole fuelling rate which is further modulated by the strength of the jet feedback.

\subsection{Evolution of the stellar component}
\label{Subsec: Stars}

Having examined the evolution of the black holes and jets, we now explore that of the stellar component. The left-hand panel of Fig.~\ref{fig: SFR} shows the time evolution of the SFR, averaged in $10$~Myr windows, and the right-hand panel shows the time evolution of the mass of newly formed stars\footnote{`Newly formed stars' refers to stars that have formed stochastically during the simulations via the processes described in Section~\ref{Sec: Additional physics}. There is a non-zero initial mass of newly formed stars due to the fact that we include in the total, the mass of stars formed during the `initial' simulations that precede our `production' runs (as described in Section~\ref{Sec: Set up}).} that have formed in the simulation.

Before exploring how the jets affect the stellar component, we first discuss the physical processes responsible for some of the general features in the evolution of the SFRs. Recall that all simulations begin as the galaxies approach first passage. The tidal torques during this encounter remove angular momentum from the gas, leading to rapid nuclear gas inflows and enhanced central densities. This, in turn, leads to the early peak in the SFR of $\sim 100 \, {\rm M_\odot \, yr^{-1}}$ and the subsequent decline that is seen in all of the simulations. A more significant peak in the SFR is seen during the final coalescence of the galaxies at $\sim 0.8$~Gyr. These two peaks in the SFR at first passage and coalescence are typical of galaxy mergers \citep[see e.g.][]{2005Springel+}, however the magnitude of the SFR and the relative heights of the peaks is highly dependent on the properties of the merging galaxies.

During coalescence, the star formation rates can reach $\sim 700 \, {\rm M_\odot \, yr^{-1}}$ in the simulation without jets. The simulation with jets, however, shows a lower peak SFR at coalescence, that only reaches $\sim 400 \, {\rm M_\odot \, yr^{-1}}$. This reduction in SFR is also reflected in the mass of stars that form, with the jet simulation forming and maintaining a lower stellar mass. It is worth reiterating the fact that jet-driven suppression of the SFR by a factor of $\sim 2$ is possible despite relatively little black hole growth (see the discussion in Section~\ref{Subsec: Jet properties}). Whilst the SFR does drop after the peak, during the coalescence, it is relatively gradual and star formation does continue during this time and the galaxies do not undergo `instantaneous quenching' as a result of the merger. If the merger were to be responsible for quenching the galaxies, then this process must, therefore, occur on longer timescales than were captured in these simulations. Furthermore, this could perhaps also indicate that jets alone do not lead to quenching during a merger, and that we do additionally need to include the effects of AGN winds and/or radiation. Alternatively, it may be that the black hole accretion rates and jet powers in these simulations are too low. Future radio observations of jetted merging galaxies with e.g. the ngVLA and SKA will help us constrain jet energetics in these systems.

It is interesting to note that the presence of jets acts to suppress the SFR throughout the majority of the merger process and does not lead to the triggering of significant star formation. This is reflected in the total stellar mass that has ever formed which is a factor of up to $1.5$ times higher in simulation without jets compared to that with jets. At this point it is worth mentioning that we do not allow star particles hosted by jet cells (i.e. those with a jet fraction greater than $10^{-3}$) to return mass, nor metals to their surroundings as we found that this often results in artificial features in the jet lobes. Doing so does, however, mean that mass can remain locked up in stars for longer. Ultimately, this means that the mass of the (evolved) stellar component in the simulation with jets is typically larger than expected, rising to $\sim 6\times10^{10}\; {\rm M_\odot}$ just before the final coalescence of the galaxies and $\sim 7\times10^{10}\; {\rm M_\odot}$ at the end of the simulation, compared to the final stellar masses of $\sim 8\times10^{10}\; {\rm M_\odot}$ found in the simulation without jets.

In each panel of Fig.~\ref{fig: SFR} the set of coloured vertical dotted lines at $\sim 0.4$~Gyr indicate the time at which the galaxies reach the apocentre of their orbit. The set of vertical lines at $0.8-0.9$~Gyr indicate the time when for the first time the SMBH separation is less than $2$ times the smoothing length of the black holes. Considering these and the timings of the peak in the SFR associated with coalescence, it is clear that the orbital progression is altered in the simulation with AGN jets. This apocentre delay may, in part, be due to increased rate of expulsion of gas from the system in the presence of AGN jets. There are, however, other factors that may also play a role such as the differences in effective resolution that arise due to the targeted refinement criteria associated with the jets \citep[see e.g.][]{2014Hayward+}.

\subsection{Properties of the large-scale outflows}
\label{Subsubsec: Outflow properties}

In this section we explore the properties and evolution of the large-scale jet-driven outflows, and quantify their multiphase nature. As discussed in Section~\ref{Subsec: Qualitative}, SN-driven galactic winds are present in all our simulations. These winds in our setup are generally unable to propagate beyond $\sim10$~kpc from the midplane and so, to focus the discussion on the jet-driven outflows, our analysis in this section is restricted to gas that lies between $20-100$~kpc above the midplane and within a cylindrical radius of $100$~kpc relative to the centre of the simulation domain.

\subsubsection{Mass inflow and outflow rates}
\label{Subsubsec: Mass inflow/outflow}

\begin{figure*}
    \centering
    \includegraphics[width=\textwidth]{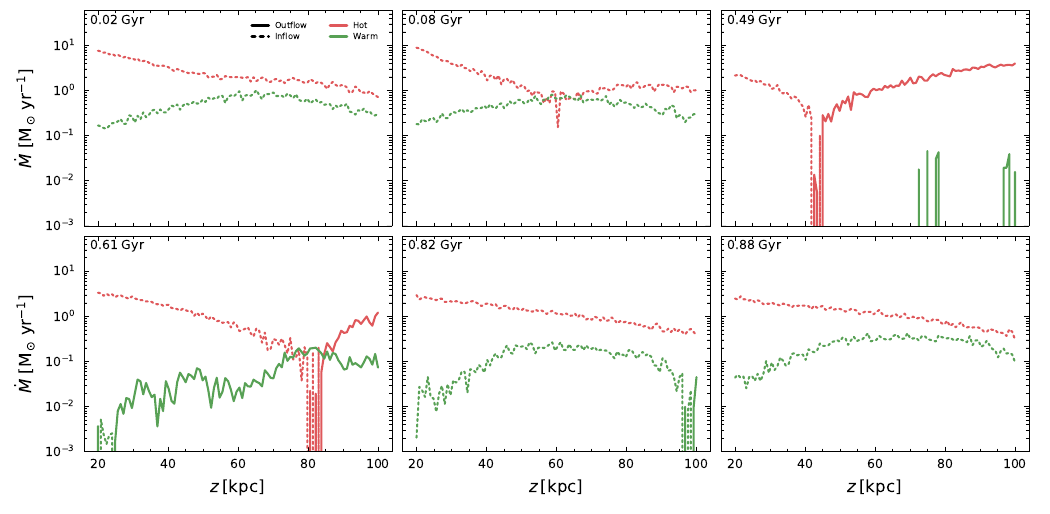}
    \caption[Time evolution of vertical mass outflow rate profiles for the simulation without jets]
    {Time evolution of the vertical mass outflow rate profiles for the simulation \textit{without} jets. Each panel shows gas between $20-100$~kpc above the midplane and within a cylindrical radius of $100$~kpc relative to the centre of the simulation domain. The time to which the profiles correspond is indicated in the top left of each panel. The mass outflow rate is split into contributions from hot ($T>5\times10^5$~K) and warm ($2\times10^4<T<5\times10^5$~K) gas (see legend in the top left panel). Note that, in this simulation, there is no cold ($T<2\times10^4$~K) gas present at these vertical distances from the midplane. Solid/dotted curves indicate a net outflow/inflow of gas.}
    \label{fig: mdot out cool nojet}
\end{figure*}
\begin{figure*}
    \centering
    \includegraphics[width=\textwidth]{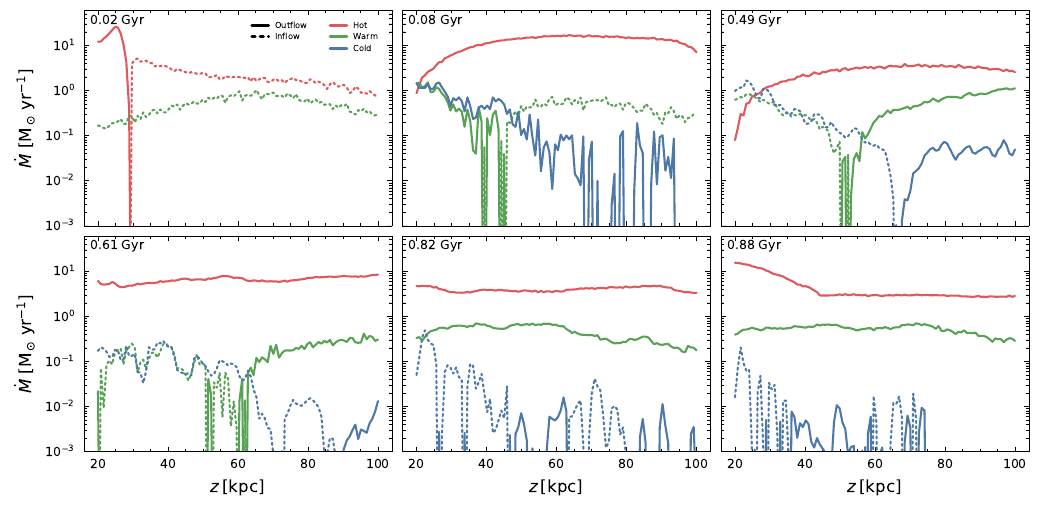}
    \caption[Time evolution of vertical mass outflow rate profiles for the simulation with jets]
    {Time evolution of the vertical mass outflow rate profiles for the simulation \textit{with} jets. Each panel shows gas between $20-100$~kpc above the midplane and within a cylindrical radius of $100$~kpc relative to the centre of the simulation domain. The time to which the profiles correspond is indicated in the top left of each panel. The mass outflow rate is split into contributions from hot ($T>5\times10^5$~K), warm ($2\times10^4<T<5\times10^5$~K) and cold ($T<2\times10^4$~K) gas (see legend in the top left panel). Solid/dotted curves indicate a net outflow/inflow of gas.}
    \label{fig: mdot out cool}
\end{figure*}

In this section, we examine the mass outflow rates that these jets drive. Since non-negligible gas motions exist in the absence of jets, we must analyse the jet-driven outflows by identifying and quantifying the ways in which the outflows differ from those found in the simulations without jets. To this end, Figs~\ref{fig: mdot out cool nojet}~and~\ref{fig: mdot out cool} show the time evolution of the vertical mass outflow rate profiles for the simulations without and with jets, respectively. In these figures, each panel shows the vertical mass outflow rate profile at times which are indicated in the top left. To examine the phase-structure of the outflow, we split the mass outflow rate into contributions from `hot' ($T>5\times10^5$~K), `warm' ($2\times10^4<T<5\times10^5$~K) and `cold' ($T<2\times10^4$~K) gas. Note that solid/dotted curves correspond to a net outflow/inflow of gas.

In the simulation without jets the gas motions at these altitudes must arise due to the merger dynamics and cooling of the hot halo. In the first and second panels of the top row of Fig.~\ref{fig: mdot out cool nojet}, gas motions are dominated by inflows, with warm and hot gas flowing back towards the midplane as the gas cools and falls into the potential well of the system. In the third panel, there is very little net motion of warm gas and, beyond $40$~kpc, all hot gas is outflowing. This behaviour comes about due to the propagation of the quasi-spherical shock that results from the first passage of the galaxies. This outflow is initially seen in the hot gas but is subsequently followed by a slower-moving warm outflow (as can be seen in the panel corresponding to $0.61$~Gyr). After the passage of this shock the gas settles into a new equilibrium, begins to cool, and net inflows resume. In the final panel, the galaxies have coalesced but the shock associated with this merger is much weaker and has not had time to propagate far enough to affect these outflow rate profiles.

\begin{figure*}
    \centering
    \includegraphics[width=\textwidth]{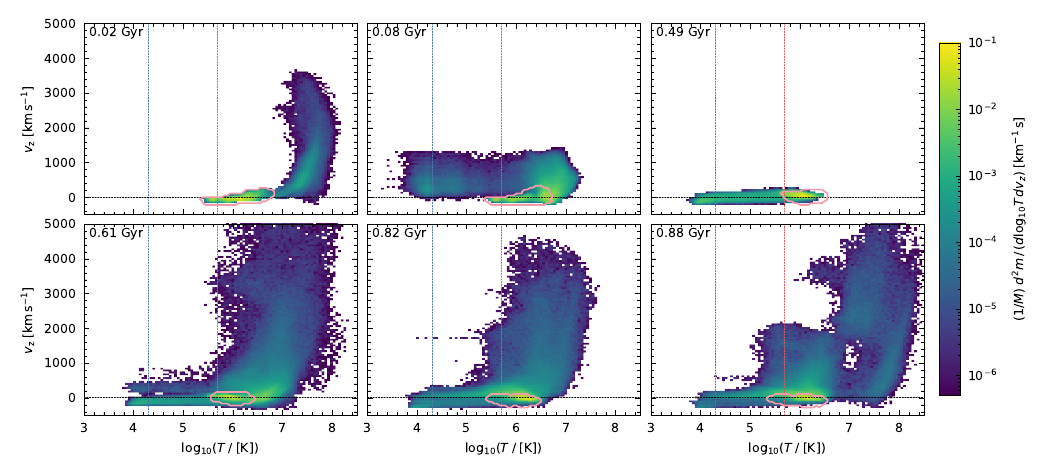}
    \caption[Joint PDFs in temperature and vertical velocity for gas in the jet-driven outflows for the simulation with jets]
    {Each panel shows, for the simulation with jets, mass-weighted, joint PDFs in $\log_{10}(T)$ and $v_z$ for gas within a cylinder of radius $100$~kpc relative to the centre of the simulation domain and with vertical extent $20$-$100$~kpc above the midplane. The time to which each PDF corresponds is indicated in the top left of the panel and is the same as those shown in Figs~\ref{fig: mdot out cool nojet}~and~\ref{fig: mdot out cool}. In each panel, the pink contours outline the region of phase-space occupied in the matching simulation without jets. The vertical blue and red dashed lines separate the temperature space into cold, warm and hot gas, using the same definitions as in Figs~\ref{fig: mdot out cool nojet}~and~\ref{fig: mdot out cool}.}
    \label{fig: T vs Vz mass}
\end{figure*}

Turning now to the simulation with jets, the early stages of a hot, jet-driven outflow can be seen in the first panel of Fig.~\ref{fig: mdot out cool}. The warm gas profile is largely identical to that shown in Fig.~\ref{fig: mdot out cool nojet}, indicating that, at these early times, the jets are largely having an effect on the hot gas. After $0.08$~Gyr, the hot outflow has already propagated out to $100$~kpc and, beyond this time, a net hot outflow of $\sim10\, {\rm M_\odot \, yr^{-1}}$ that is fairly constant in height above the midplane persists throughout the remainder of the simulation\footnote{Note, however, that when integrated, the mass loss rate for the simulation with jets corresponds to $< 1\%$ of the gas disc mass per Gyr.}. In the final panel (bottom right), however, there is evidence of an enhancement in this hot outflow rate which is due to the powerful jet outburst associated with the final coalescence of the galaxies, as seen in the bottom row of Fig.~\ref{fig: Various quantites}.

Whilst, initially, there is a net inflow of warm gas, after $0.08$~Gyr the warm gas in the inner regions becomes outflowing due to the action of the jets. By $0.49$~Gyr however, the jets have become relatively quiescent and the inner regions have returned to a state of net inflow. The fact that the warm outflow takes longer to propagate than the hot component indicates that it is moving more slowly. Indeed, this is the case, and we will discuss the dynamics of the gas shortly. When the jet activity increases again, net outflows of warm gas can be supported throughout the entire domain considered (see the final two panels).

One of the most obvious differences between Figs~\ref{fig: mdot out cool nojet}~and~\ref{fig: mdot out cool} is the presence of cold gas in the outflows of the latter. In the simulation without jets, all non-negligible gas motions occur in the hot and warm component meaning that the cold component in Fig.~\ref{fig: mdot out cool} must be attributed to the action of the jets. Processes that are likely responsible for the development of a cold phase are the condensation of the warm phase associated with the jet-driven outflows, the entrainment of SN-driven wind material and the direct expulsion of gas from the galactic discs, as discussed in Section~\ref{Subsec: Qualitative}.

After $0.08$~Gyr, a significant cold gas component has already formed which, at this time, has a net outflow rate at all heights above the galactic disc, out to $\sim95$~kpc. The mass of cold gas in this region can be as high as $\sim 10^8 \; {\rm M_\odot}$, corresponding to $\sim1$ per cent of the initial mass of gas in the galactic discs. As in the case of the warm component, when the jets are less active (see the third panel in the top row of Fig.~\ref{fig: mdot out cool}) they are not able to maintain a net outflow and the inner regions become inflowing. At later times, there is evidence of both inflowing and outflowing cold gas at a range of different heights above the midplane. This highlights the fact that the cold gas does not exist as a `monolithic' entity, but rather in clumps, clouds and streams (see e.g. Fig.~\ref{fig: Various quantites}). 

In this section we have shown that significant amounts of cold outflowing and inflowing gas can be produced due to the action of the jets, thus enhancing the multiphase nature of the surrounding CGM and halo gas. It is also worth mentioning that the typical mass outflow rates in the cold gas (and even more so for the hot gas) are comparable or even higher than the mass inflow rate onto the $\alpha$-disc (shown in Fig.~\ref{fig: subgrid multiple}), highlighting the efficiency by which our simulated jets are able to launch these large scale outflows.

\subsubsection{Outflow dynamics}
\label{Subsubsec: Outflow dynamics}

\begin{figure*}
    \centering
    \includegraphics[width=0.98\textwidth]{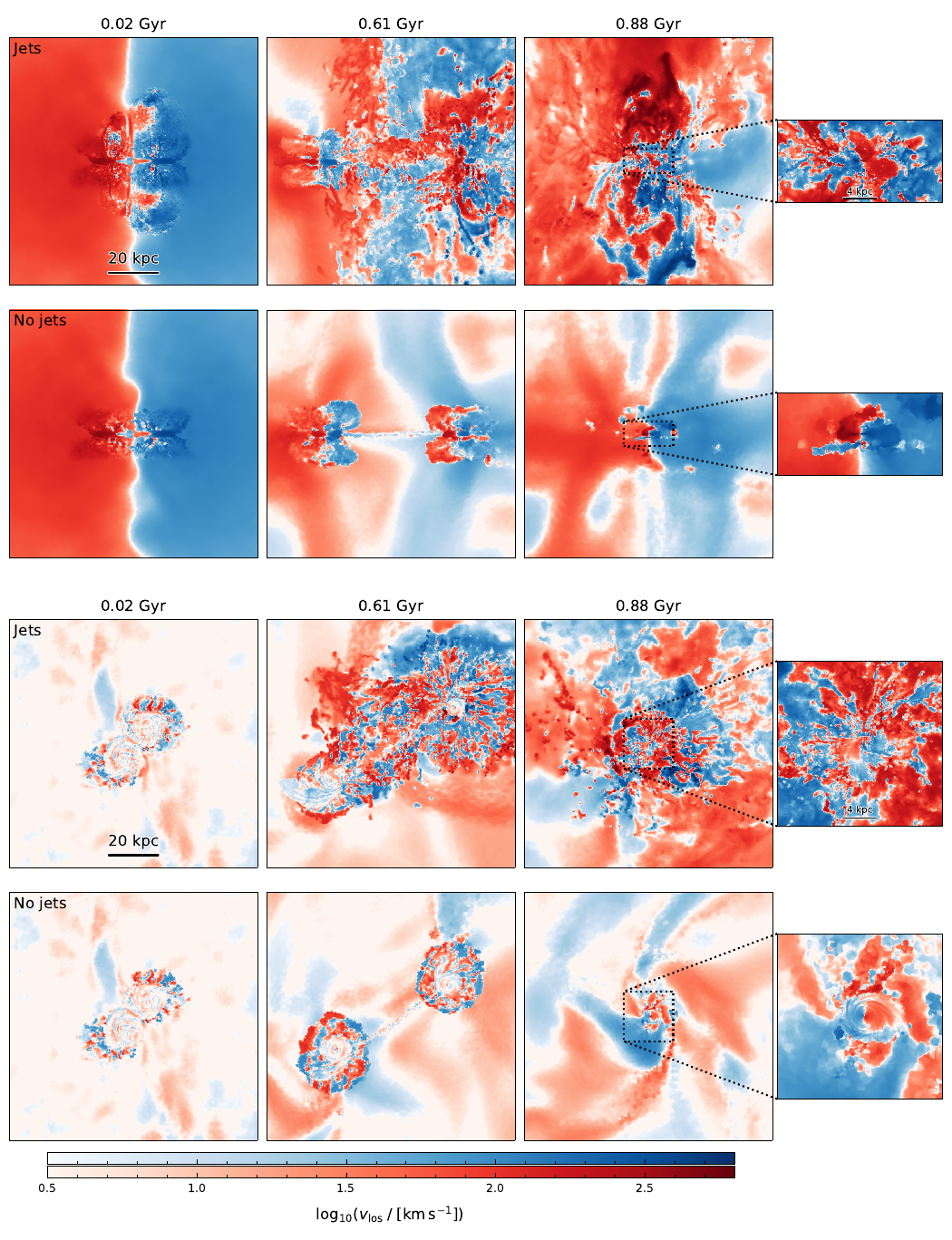}
    \vspace{-0.5cm}
    \caption[Emission-weighted LoS velocity maps for simulations with and without jets]
    {Emission-weighted LoS velocity maps. The projections in the first two rows are edge-on and those in the third and fourth rows are face-on. The first and third rows correspond to the simulation with jets while the second and fourth rows show the simulation without jets. The first three columns correspond to projections of gas within a volume of $100\times 100\times 100$~kpc and the time at which the projection was made increases from left to right and is indicated in the column headings. The fourth column shows a zoom-in of the third, with dimensions $20\times20\times10$~kpc and $20\times10\times20$~kpc for the face-on and edge-on projections, respectively.}
    \label{fig: los vel all}
\end{figure*}

\begin{figure*}
    \centering
    \includegraphics[width=0.98\textwidth]{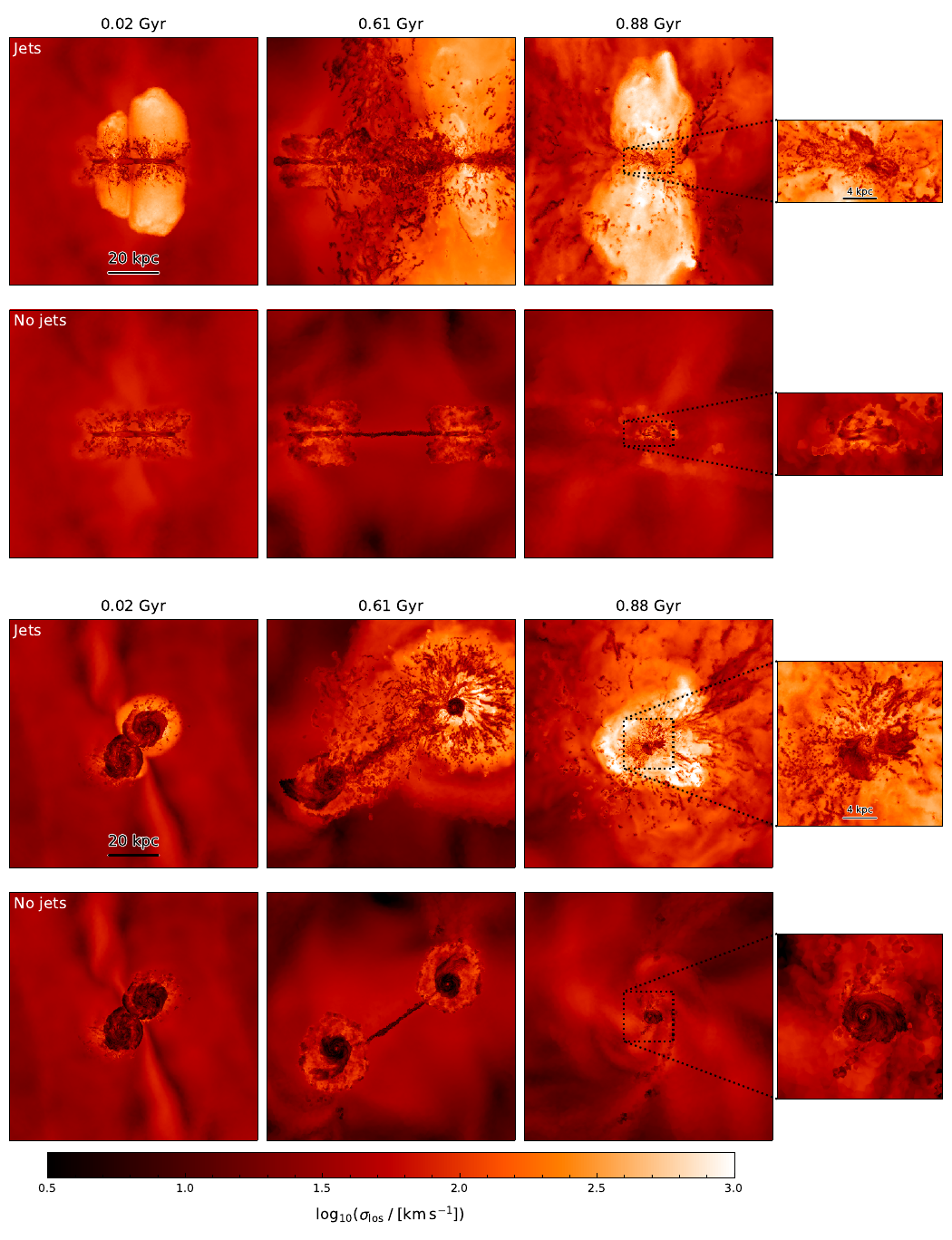}
    \vspace{-0.5cm}
    \caption[Emission-weighted LoS velocity dispersion maps for simulations with and without jets]
    {Emission-weighted LoS velocity dispersion maps. The projections in the first two rows are edge-on and those in the third and fourth rows are face-on. The first and third rows correspond to the simulation with jets while the second and fourth rows show the simulation without jets. The first three columns correspond to projections of gas within a volume of $100\times 100\times 100$~kpc and the time at which the projection was made increases from left to right and is indicated in the column headings. The fourth column shows a zoom-in of the third, with dimensions $20\times20\times10$~kpc and $20\times10\times20$~kpc for the face-on and edge-on projections, respectively.}
    \label{fig: vdisp all}
\end{figure*}

In this section we examine the dynamics of these jet-driven gas motions with the aid of Fig.~\ref{fig: T vs Vz mass}, which shows mass-weighted, joint probability density functions (PDFs) in $\log_{10}(T)$, and vertical velocity, $v_z$, for the same volume of gas used to calculate the outflow profiles in Figs~\ref{fig: mdot out cool nojet}~and~\ref{fig: mdot out cool}. The time to which each PDF corresponds is indicated in the top-left of the panel and is the same as those shown in Figs~\ref{fig: mdot out cool nojet}~and~\ref{fig: mdot out cool}. Additionally, the vertical blue and red dashed lines separate the temperature space into cold, warm and hot gas, defined in the same way as in Section~\ref{Subsubsec: Mass inflow/outflow}. 

The pink contour in each panel outlines the region of phase-space occupied in the simulation \textit{without jets}. From this, it is clear that in the absence of jets, the majority of the gas is hot and relatively slow moving, with maximum velocities typically not exceeding $500 \, {\rm km \, s^{-1}}$. There is also a non-negligible inflowing component in both the warm and hot gas, consistent with the discussion in Section~\ref{Subsubsec: Mass inflow/outflow}. 

The first panel in the top row corresponds to just after the jets have been launched and the hot, outflowing component is clearly visible, reaching temperatures of up to $10^8$~K and velocities approaching $4000\, {\rm km\,s^{-1}}$. By the time $0.08$~Gyr have elapsed, the velocities of the gas, while significantly enhanced above what would be found if the jets were not present, no longer reach such extreme speeds. At this time, the hot gas has begun to cool and has led to the formation of an outflow of warm gas and an increase in the mass of the cold phase. Additionally, some of this cold and warm gas is now inflowing, albeit at comparatively low velocities.

After $0.49$~Gyr, the gas has slowed considerably and a non-negligible fraction of the cold and warm phases are inflowing, facilitating the net warm and cold inflows seen in the inner regions in the third panel of Fig.~\ref{fig: mdot out cool}. When the jets become active again, as is the case in the bottom row of Fig.~\ref{fig: T vs Vz mass}, the hot, fast, outflowing phase is replenished, with gas able to reach even more extreme velocities. At $0.61$~Gyr, the cold and warm inflowing components remain and the outflow velocities are somewhat higher, but still significantly slower than those observed in the hot gas. At $0.82$~Gyr, the hot outflow has cooled somewhat and has begun to repopulate the warm phase with gas that has velocities reaching $\sim2500\, {\rm km\,s^{-1}}$. 

In the final panel, the gas associated with the jet outburst that populated the hot, fast phase in the previous two panels has cooled and slowed and, now, largely consists of gas with temperatures in the range $10^5 - 10^7$~K with velocities that reach $2500\, {\rm km\,s^{-1}}$. The powerful jet outburst associated with the merger is also clearly visible as it replenishes the hot phase with gas of temperatures and velocities that can exceed $10^8$~K and $5000\, {\rm km\,s^{-1}}$, respectively.

After the jets launch, the total mass of cold gas that lies beyond $20$~kpc from the midplane rapidly rises to $\sim10^7 \, {\rm M_\odot}$ and can even exceed $10^8 \, {\rm M_\odot}$ at times. Interestingly, the cold outflows can also have very high vertical velocities that reach $\sim 2500 \, {\rm km\,s^{-1}}$.

\subsection{Line-of-sight velocities and velocity dispersions}
In the previous section we explored the velocity structure of the gas above the galactic discs. Here, we extend this analysis and examine spatially resolved maps of the line-of-sight (LoS) velocity and velocity dispersion in the merging system.

Even in the absence of jets, the kinematics of the gas in a merging system are highly complex. Non-negligible gas motions arise due to the orbital motion of the galaxies, the rotation of the gas discs and the fact that the gas in the halo is not static. To understand how the presence of jets affects the gas kinematics we must, therefore, analyse the velocity structure of the simulations with jets through comparison to that of the simulations without jets. Ultimately this allows us to isolate the features of the velocity field that can be attributed solely to the action of the jets.

Figs~\ref{fig: los vel all}~and~\ref{fig: vdisp all} show, respectively, emission-weighted LoS velocity and LoS velocity dispersion maps. The first two rows of Figs~\ref{fig: los vel all}~and~\ref{fig: vdisp all}, show edge-on projections and the third and fourth rows show face-on projections. The projections in the first three columns have dimensions of $100\times 100\times 100$~kpc, whilst the fourth column shows a zoom-in of the third, with dimensions $20\times20\times10$~kpc and $20\times10\times20$~kpc for the face-on and edge-on projections, respectively. The first and third rows correspond to the simulation with jets while the second and fourth rows show the simulations without jets. The time at which the projection was made increases from left to right and is indicated in the column headings. Note that these times are the same as those shown in Fig.~\ref{fig: Various quantites}.

At early times, in both the jet and no-jet cases, Fig.~\ref{fig: los vel all} shows that the most prominent feature in the LoS velocity field is orbital motion of the galaxies. As well as this, the rotation of the galactic discs and the outflows and inflows associated with the galactic winds are clearly imprinted in the edge-on and face-on projections, respectively. In the edge-on projections, it is possible to discern the jets, although their impact on the structure of the velocity field is relatively moderate at this time. The effects of the jet on the velocity dispersion (Fig.~\ref{fig: vdisp all}), however, are much more obvious. The jets have inflated cocoons of turbulent gas that has velocity dispersions reaching several $100 \; {\rm km \, s^{-1}}$. These jet-driven outflows with high velocity dispersions are clearly distinct from the outflows associated with the galactic winds which, we can see, typically exhibit much lower velocity dispersions, and provide a good way for observations with spatially resolved kinematics to distinguish these two different types of outflows. 

After $0.61$~Gyr (the middle columns of Figs~\ref{fig: los vel all}~and~\ref{fig: vdisp all}) the jets are now clearly having an impact on the gas velocity, with significant enhancements seen in both the vertical and the lateral components of the velocity. This is particularly evident when considering the jet launched from the galaxy on the right-hand side of these projections which, as has been discussed in previous sections, is more powerful at this time. As well as having a greater impact on the velocity field, this jet is also clearly causing significant enhancement of the velocity dispersions, indicative of highly turbulent flows. The jet on the left-hand side, however, is relatively quiescent at this time and the velocities and velocity dispersions of the gas in the vicinity of this galaxy are lower. Nevertheless, both jets are associated with velocities and velocity dispersions that are higher than those found in the simulation without jets, in which the highest velocities are largely confined to the galaxy discs and the winds, which typically do not extend further than $10$~kpc from the orbital plane.

One other feature to highlight, is the fact that the velocity distribution of the gas in the jet-driven outflows has multiple components. This comes about due to the fact that, as discussed in Section~\ref{Subsec: Jet properties}, the power of the jets and, therefore, the velocity of the gas at the base of the jet, can be significantly variable. Additionally, as the jets propagate away from the midplane, they interact with the galactic winds. Altogether this acts to disrupt the jet-lobes and enhances the levels of turbulence in the resulting outflows, which is clearly seen in Fig.~\ref{fig: vdisp all}.

As has been discussed in previous sections, at $0.61$~Gyr gas is condensing out of the jet-driven outflows (see Fig.~\ref{fig: Various quantites}) and falling back towards the orbital plane of the galaxies. From Fig.~\ref{fig: vdisp all} it is clear that this cooling gas typically has lower velocity dispersions which are comparable to those found in the galactic winds than those of the newly launched jet material.

At later times (third column of Figs~\ref{fig: los vel all}~and~\ref{fig: vdisp all}), as the galaxies coalesce, the fast, bipolar outflow associated with the powerful, post-merger jets is clearly visible in the LoS velocity map, with velocities exceeding $1000 \; {\rm km \, s^{-1}}$ seen in both the edge-on and face-on maps. Due to the fact the more powerful jet is significantly inclined to the vertical at this time (see Fig.~\ref{fig: subgrid multiple}), both the redshifted and the blueshifted components of the outflow can be seen in the face-on map. The velocity dispersions also exhibit significant enhancement, and can exceed velocities of $1000 \; {\rm km \, s^{-1}}$. This is in contrast to the run without jets where the velocities and velocity dispersions remain moderate, despite the violent motions associated with the coalescence of the galaxies. These differences between the mergers with and without jets can also be seen in the inset projections, shown in the final columns of Figs~\ref{fig: los vel all}~and~\ref{fig: vdisp all}, which highlight the considerable impact that jets can have on the properties of the merger remnant.
\begin{figure}
    \centering
    \includegraphics[width=\columnwidth]{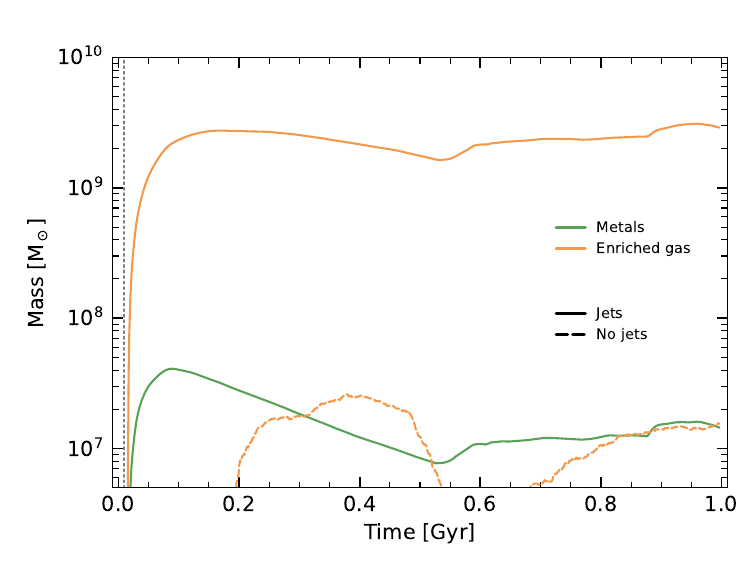}
    \caption[The mass of enriched gas and the total metal mass in the outflow]
    {The mass of enriched ($Z > {0.1\,Z_\odot}$) gas (orange curve) and the total metal mass (green curve) in the gas that lies between $20$ and $100$~kpc above the midplane and within $100$~kpc of the centre of the simulation domain. The solid/dashed curves correspond to the simulations with/without jets.}
    \label{fig: enrichment}
\end{figure}

\subsection{Metal enrichment}

In this section, we investigate the extent to which the jets are able to influence the metallicity of the gas in the halo. Fig.~\ref{fig: enrichment}, shows the mass of enriched ($Z >0.1\,{\rm Z_\odot}$) gas (orange curve) and the total metal mass (green curve) in the gas that lies between $20$ and $100$~kpc above the midplane and within $100$~kpc of the centre of the simulation domain (this is the same region as considered in Sections~\ref{Subsubsec: Mass inflow/outflow}~and~\ref{Subsubsec: Outflow dynamics}). The solid/dashed curves corresponding to the simulations with/without jets.

It is clear that, after $\sim0.1$~Gyr there is a non-negligible mass of enriched gas in this region in the simulations with jets. A mass of at least $2\times10^9 \; {\rm M_\odot}$ of enriched gas is then maintained in the halo throughout the rest of the simulation. This observation is consistent with our discussion in Section~\ref{Subsec: Qualitative}, where the projections in Fig.~\ref{fig: Various quantites} indicated that the metallicity of the halo in the simulations with jets is enhanced relative to that in the simulation without jets. Indeed, Fig.~\ref{fig: enrichment} shows that the simulation without jets have very little enriched gas present in this region, above the midplane, with the mass of enriched gas only briefly exceeding $10^7 \; {\rm M_\odot}$. Examining radial metallicity profiles (not shown in this paper) we find that, with jets, the average metallicity is above $0.1\,{\rm Z_\odot}$ out to $\sim 75$~kpc at first passage and then beyond $150$~kpc post-merger. When jets are not present, however, the average metallicity remains above $0.1\,{\rm Z_\odot}$ at all times only within a sphere of radius $\sim 20$~kpc.

The mass of metals in this region is non-negligible at all times after $\sim0.1$~Gyr in the simulations with jets and largely stays above $10^7 \; {\rm M_\odot}$. On the other hand, the metal mass for the simulations without jets is not shown in Fig.~\ref{fig: enrichment} as we found it to be insignificant at all times. Focusing on the simulations with jets and comparing the time evolution of the mass of enriched gas and metals to that of the jet power, shown in Fig.~\ref{fig: subgrid multiple}, one can tentatively associate times at which the metal content and the mass of enriched gas increases with peaks in the jet power (with a slight time-delay due to the time it takes to `communicate' changes in jet power to the halo gas).

Altogether, this analysis highlights the fact that jets can be very efficient at drawing enriched material up out of the galaxy and into the halo. In fact, it is likely that the jets would be even more efficient at enriching the halo gas due to the fact that, as discussed in Section~\ref{Subsec: Stars}, mass and metal return from stars in the jet lobes is purposefully suppressed in our simulations.

%%%%%%%%%%%%%%%%%%%%%%%%%%%%%%%%%%%%%%%%%%%%%%%%%%%%%%%%%%%%%%%%%%%%%%%%%%%%%%%%%%%%%%

\section{Discussion: Black holes in galaxy mergers}
\label{Sec: Discussion}

Previous works that have investigated galaxy mergers using simulations have shown that the inclusion (or lack thereof) of wide range of physical processes and components such as: galactic winds, black hole accretion and feedback and the presence of a hot, gaseous halo or a stellar bulge can have significant and varied impacts on the progression and outcome of the merger. In this section we first discuss the ways in which the effects of kinetic AGN jets on the mergers differs in comparison to other merger simulations and outline some of the limitations of our simulations.

\subsection{The effects of black hole jets in galaxy mergers}

Many idealised merger simulations that model black hole feedback processes do so by injecting thermal energy into the gas cells local to the black holes \citep{2005Springel+,2005DiMatteo+,2006Robertson+,2009Johansson+,2014Hayward+,2016Gabor+}. Some also consider injecting a kinetic wind \citep{2014Barai+, 2020Costa+,2020Torrey+}. Investigations of AGN feedback from black holes in galaxy mergers have also been carried out within the context of large-volume cosmological simulations \citep[see e.g.][]{2019RodriguezMontero+,2020Hani+,2021Quai+} and also in cosmological zoom simulations \citep[see e.g.][]{2016Sparre+,2021Whittingham+}. The effects of kinetic AGN jet feedback in merging galaxies (as investigated in this work), however, remains largely unexplored.

When the galaxies in our simulations pass each other for the first time, they experience strong tidal forces which lead to the formation of extended tidal tails and bridges of gas and stars. As the galaxies approach for the second time during the final coalescence, the tidal torques are even more extreme, ultimately resulting in the formation of a compact dense merger remnant. This general picture of the progression of the merger and the qualitative morphology of the two galaxies in our simulations are consistent with the majority of existing literature \citep[see e.g.][]{2005Springel+,2006Cox+b,2009Johansson+,2010Teyssier+,2011Moster+,2013Hopkins+,2014Hayward+} and is, therefore, largely unchanged by the presence of the jets.

In general, the behaviour of the SFR (see Section~\ref{Subsec: Stars}) in our simulations is also qualitatively similar to those found in other merger studies \citep[provided that the galaxies in question have stellar bulges; see e.g.][]{2005Springel+,2009Johansson+, 2011Moster+}. The magnitude of the SFR in our simulations is only moderately suppressed when jets are present (see Section~\ref{Subsec: Stars}). We do not, however, see evidence for rapid quenching as a result of the final coalescence of the galaxies. This is in contrast to some works who find that extreme tidal torques during the final coalescence drive significant inflows, enhancing the black hole accretion rate and powering the AGN which results in star formation being shut off and black hole growth stalling \citep[see e.g. seminal papers by][]{2005DiMatteo+,2005Springel+}.

One of the reasons for this difference may be due to the fact that the AGN feedback models used in such simulations are typically intended to reproduce the effects of the `quasar mode'. This is often implemented via an isotropic injection of energy, whereas the jets in our simulations are highly collimated, meaning that a greater fraction of the injected energy is likely deposited in the halo, rather than impacting the galactic disc. The jets, therefore, typically do not clear out significant quantities of gas from the centre and likely have a comparatively smaller effect on the outer disc when compared to an isotropic energy injection, ultimately making it harder for jets to quench star formation and suppress the growth of the black holes. Detailed numerical simulations of a variety of AGN feedback mechanisms (e.g. energy-, radiation-pressure-driven outflows and jets) within a spatially resolved ISM will be needed to understand which physical processes lead to galaxy quenching.

The reduction of the SFR that we find occurs when jets are present is in general agreement with studies that suggest that jets may prevent catastrophic cooling in clusters and regulate the gas accretion rate and, therefore, the star formation rate \citep[see e.g.][]{2011Gaspari+,2016YangReynolds,2020Prasad+}. But other studies find that jets may have a positive feedback effect, resulting in enhanced star formation rates \citep{2012Gaibler+,2017Fragile+,2018Mukherjee+}. It is likely, however, that jets are capable of having both a positive and a negative effect on the star formation rate \citep[see e.g.][]{2021Mandal+}, depending on the properties of the jet as well as the local- and wider-environment. To shed further light on the effects of jets on the SFR during a galaxy merger, analogous simulations would need to be carried out with a more accurate model for star formation and stellar feedback, including a resolved ISM, as well as probing a wider range of merging systems.

Additionally, simulations have shown that jets may act to change to location of star formation \citep{2012NayakshinZubovas,2017ZubovasBourne} with observations indicating that star formation in AGN-driven outflows may well be occurring \citep[see e.g.][]{2017Maiolino+}. So, while the total SFR may be reduced by AGN, the location of the star formation may be changed, with enhancements seen in some areas and reductions seen in others.

Another point worth noting is that our black hole accretion prescription differs from the commonly-used Bondi-Hoyle-like prescriptions. As discussed in Section~\ref{Subsec: Jet properties}, the \textit{inflow} rates onto the sub-grid $\alpha$-disc that we measure in our simulations can be rather sporadic due to the fact that we only allow gas to reach the sub-grid $\alpha$-disc if it is radially inflowing \textit{and} able to circularise at a radius smaller that the outer edge of the $\alpha$-disc. With this accretion prescription, we find that accretion rate due to the coalescence of galaxies is not as significant as those in \citet{2005DiMatteo+,2005Springel+} and is comparatively short lived. This likely also contributes to the fact that the AGN outburst associated with coalescence that we observe is less effective at quenching the galaxies and suppressing black hole growth. We do, however, find that, before the coalescence, the magnitude of the inflow rates in our simulations are largely similar to those measured in analogous simulations \citep{2005Springel+}. This highlights a very important open question in the field, and more studies focusing on accurate and realistic modelling of black hole accretion are needed to understand how gas is delivered from large scales to the actual accretion discs. It is worth emphasising that Bondi-like accretion flows are radiatively inefficient. Therefore, to explain the quasar phenomena, ultimately, simulations modelling accretion discs, as attempted here, are required. 

Whilst not shown explicitly in this paper, we find that the black hole mass growth throughout the simulation is rather moderate. This is even true during the final coalescence of the galaxies where we do find somewhat elevated accretion rates. This can be attributed to the fact that, in our jet feedback model, the black hole spin energy (and, therefore, its mass) is the source of energy powering the jet, along with the accretion flow (this was discussed at length in \citet{2022Talbot+}). This is in contrast to many other works that focus on black holes in merger scenarios, where the black hole growth typically traces the accretion rate.

In Section~\ref{Subsec: BH feeding} we saw that the launching of the jets can induce the condensation of hot gas out in the halo. This gas can then fall back onto the galaxies and provide additional fuel for the black hole. Such processes have been proposed and investigated in works such as \citet{2013Gaspari+,2015Gaspari+}, where this process is termed `Chaotic Cold Accretion' \citep[see also][]{2012McCourt+,2015Prasad+,2019Wang+}. Such processes clearly have the potential to be an important source of fuel for black holes, both within merger scenarios and otherwise. But, as highlighted here, `secular' processes, merger-induced torques and jet-driven backflows \citep{2010Antonuccio-DeloguSilk,2014Cielo+,2017BourneSijacki} also lead to fresh gas inflows onto the $\alpha$-discs. 

\subsection{Limitations of the simulations}

For an extensive discussion of the key assumptions and limitations of our black hole accretion and jet feedback model, we refer the reader to \citet{2021Talbot+} and \citet{2022Talbot+}. It is, however, worth reiterating a few key points that are particularly relevant to the work presented in this paper. Firstly, we use GRMHD simulation data to parameterise the magnetic flux on the black hole horizon and the data we use corresponds to that which would be found when the accretion disc is in a magnetically arrested (MAD) state. This means that, for a given accretion rate, the powers of the jets are likely on the high end of what would be expected in reality. In addition, the power of the jet will depend on the choice of initial black hole spin, for which we only consider one initial, relatively high value ($a_0 = 0.7$) in this work, which lies within the range of observationally determined black hole spins \citep[see][]{2021Reynolds}. Our model also assumes that the sub-grid accretion disc is geometrically thin, following the Shakura-Sunyaev solution \citep{1973ShakuraSunyaev}. A more complete picture would require the incorporation of accretion prescriptions to describe the gas flow when the black hole is surrounded by a thick or possibly `truncated' accretion disc.

Whilst the mass of stars particles that are present in the outflow in the runs with jets is similar to those without jets, we do find that in the jet simulations the mass of star forming gas in the outflows is higher, highlighting the potential for jet-driven outflows to entrain and draw up star forming gas from the disc, as well as induce the formation of stars in the outflows themselves. To properly assess the viability and effects of star formation in the jet-driven outflows, however, a better model of star formation and feedback would be required, which explicitly models the multiphase structure of the ISM. Such a model would also be required to properly assess the effects of the jets on the stars in the disc and remnant, as well as allowing for a more accurate estimate of the black hole accretion rate.

It is also worth highlighting, again, the highly idealised nature of the setup explored in this work. Future simulation work should include the of the wider cosmological environment, so as to properly capture the effects of cosmic gas inflows which, at this redshift, should significant. We have, however, attempted to somewhat mitigate some of these effects through our choice of eEOS parameters (see Section~\ref{Subsec: eEOS}) and by modelling the hot halo gas. Future work that aims to assess the impact of jets on mergers should also probe a wider range of orbital configurations, mass ratios and black hole spin directions and magnitudes.

Finally, with regard to physical processes, these simulations do not include the effects of magnetic fields. The inclusion of MHD effects in future simulations will be crucial if we are to make accurate predictions for the radio emission and, indeed, ensure that the simulations accurately capture the gas dynamics and properties of the discs in the merger \citep[see e.g.][]{2021Whittingham+}.

%%%%%%%%%%%%%%%%%%%%%%%%%%%%%%%%%%%%%%%%%%%%%%%%%%%%%%%%%%%%%%%%%%%%%%%%%%%%%%%%%%%%%%

\section{Conclusions}
\label{Sec: Conclusions}
In this paper, we presented simulations of AGN jets in the context of a major merger of two gas-rich galaxies, such as would be found at cosmic noon. These simulations follow the progression of the merger through first passage and up to the final coalescence, modelling the black holes at the centres of both of the merging galaxies using our prescription for black hole accretion via an $\alpha$-disc and feedback in the form of a spin-driven jet. The analysis of the simulations, presented in this work, focused on exploring the fuelling of black holes and self-regulation of AGN jets as they are subject to these extreme merger environments, as well as how the presence of AGN jets affects the SFR, outflows, galaxy kinematics and CGM metal enrichment. Analysis of simulations such as these will play a central role in making precise predictions for multimessenger investigations of dual radio-AGN, which next-generation observational facilities such as LISA, Athena and SKA will make possible. The work we have presented here is a first step in this direction.

Our main results are:
\begin{itemize}
    \item Jets launched by black holes at the centre of galaxies that are undergoing a major merger are capable of driving large-scale, multiphase outflows, whose kinematic is complex and characterised by large velocity dispersions.
    \item Specifically, the jets lead to generation of significant quantities of cold gas out to distances of $\sim100$~kpc by entraining and drawing-up cold gas from the galactic discs, and also by promoting the formation of thermal instabilities in the hot halo gas.
    \item The velocities of the hot gas in the outflow can exceed $5000 \, {\rm km \, s^{-1}}$ while the warm and cold components are typically slower, but still reaching velocities of order $\sim 2500 \, {\rm km \, s^{-1}}$.
    \item The gas in the outflows can, eventually, decelerate, cool and fall back down towards the orbital plane of the galaxies. This inflowing, cool gas can provide a rich source of fuel for the black hole if it falls back onto the central regions of a galaxy, ultimately resulting in further episodes of intense jet activity.
    \item In fact, in these merger scenarios, the black hole feeding is mediated by the interplay between four distinct processes: (i) secular inflows of gas from the galactic discs, (ii) the funneling of low angular momentum gas towards the black hole by small-scale backflows, (iii) the infall of cold gas that has cooled out of the hot, jet-driven outflow and (iv) extreme merger torques driving inflows towards the centre.    
    \item The jets associated with black holes that are primarily fed by the infalling gas (rather than by gas accreted from the galactic disc) can be torqued significantly due to the fact that the infalling gas that is accreted does not necessarily have angular momentum direction perpendicular to the plane of the galactic discs.
    \item AGN jets are able to moderately suppress star formation at all times during the merger and can lower the peak SFR attained during the final coalescence of the galaxies by a factor of $\sim 2$, but they do not lead to rapid quenching of galaxies. It remains to be understood if alternative AGN fuelling or feedback are needed for rapid shutdown of star formation.
\end{itemize}

%%%%%%%%%%%%%%%%%%%%%%%%%%%%%%%%%%%%%%%%%%%%%

\section*{Acknowledgements}
We would like to thank the referee for their very thoughtful and constructive report. We would also like to thank Sandro Tacchella, Roberto Maiolino and Martin Haehnelt for useful discussions and helpful comments during the development of this manuscript. RYT, DS and MAB acknowledge the support from the ERC starting grant 638707 `Black holes and their host galaxies: co-evolution across cosmic time' and the STFC. This work was performed using the Cambridge Service for Data Driven Discovery (CSD3), part of which is operated by the University of Cambridge Research Computing on behalf of the STFC DiRAC HPC Facility (www.dirac.ac.uk). The DiRAC component of CSD3 was funded by BEIS capital funding via STFC capital grants ST/P002307/1 and ST/R002452/1 and STFC operations grant ST/R00689X/1. DiRAC is part of the National e-Infrastructure. This work used the DiRAC@Durham facility managed by the Institute for Computational Cosmology on behalf of the STFC DiRAC HPC Facility (www.dirac.ac.uk). The equipment was funded by BEIS capital funding via STFC capital grants ST/P002293/1 and ST/R002371/1, Durham University and STFC operations grant ST/R000832/1. DiRAC is part of the National e-Infrastructure. This work made significant use of the NumPy \citep{2020Harris+}, SciPy \citep{2020Virtanen+}, and Matplotlib \citep{2007Hunter} Python packages.

\section*{Data availability}
The data underlying this article will be shared upon request to the corresponding author.



\bibliographystyle{mnras}
\bibliography{references} % if your bibtex file is called example.bib

%\begin{thebibliography}{99}
%\end{thebibliography}

%%%%%%%%%%%%%%%%%%%%%%%%%%%%%%%%%%%%%%%%%%%%%%%%%%

%%%%%%%%%%%%%%%%% APPENDICES %%%%%%%%%%%%%%%%%%%%%

% Don't change these lines
\bsp	% typesetting comment
\label{lastpage}
\end{document}